\documentclass[a4paper,11pt]{article}
\pdfoutput=1
\usepackage{jheppub}
\usepackage{cancel}
\usepackage{subcaption}
\usepackage{orcidlink}
\usepackage{mathrsfs}
\usepackage{amsmath}   
\usepackage{feynmp-auto,expdlist}
\usepackage{float}

 \newcommand{\beq}[1]{\begin{equation}\label{#1}}
 \newcommand{\eeq}{\end{equation}}
 \newcommand{\bea}[1]{\begin{eqnarray}\label{#1}}
 \newcommand{\eea}{\end{eqnarray}}

\title{Particle Production and Krylov Complexity of Circular Strings Near Black Hole Horizons}
 \author{Ai-chen Li $^{a,b,c}$,\orcidlink{0000-0002-5188-2707}}
 \affiliation{\it ${}^a$ Beijing University of Technology, Beijing 100124, China}
 \affiliation[b]{Departament de F\'{i}sica Qu\`{a}ntica i Astrof\'{i}sica, Institut de Ci\`{e}ncies del Cosmos (ICCUB), Universitat de Barcelona, Mart\'{i} i Franqu\`{e}s 1, E-08028 Barcelona, Spain} 
\affiliation{\it ${}^c$ School of Science, Guangxi University of Science and Technology, 545026 Liuzhou, China}

\emailAdd{alexkenlee@163.com}

\abstract{For an infalling circular string, we study particle production, Krylov complexity, Lanczos coefficients, and operator growth induced by quantum fluctuations. Using canonical quantization in the squeezed-state formalism, we show that significant particle production arises only in the radial sector as the string approaches the black hole horizon, while angular modes remain weakly excited. Exploiting the equivalence between particle number and Krylov complexity for two-mode states, we find that nontrivial complexity scaling emerges only in the near-horizon, effectively thermalized regime, where the state approaches a thermofield double form. In this limit, the particle number exhibits a polynomial dependence on the initial position of the probe string. We further identify a linear dependence of the operator growth rate on the initial position of the probe string, suggesting a universal scaling behavior of operator growth and providing support for the complexity–volume correspondence.}

\begin{document} 
\maketitle
\flushbottom

\section{Introduction}

The dynamics of circular strings in black hole spacetimes provide a powerful tool for probing both classical and quantum aspects of gravitational physics \cite{Susskind:1993ki,Mezhlumian:1994pe,Frolov:2000kx,Silverstein:2014yza,Dodelson:2015toa,Stuchlik:2012ry,Garriga:1991ts,DeVega:1992xc,Vilenkin:1991zk}. In this work, one of our key motivations is to explore a possible connection between Krylov complexity and the holographic “Complexity = Volume” conjecture \cite{Susskind:2014rva}. Krylov complexity was originally introduced as a quantitative measure of operator growth in quantum systems \cite{Hornedal:2022pkc,Caputa:2021ori}, and has since been extended to a wide range of physical settings, including quantum field theories in flat spacetime \cite{Adhikari:2022whf,Caputa:2022eye,Banerjee:2022ime,He:2022ryk,Camargo:2022rnt,Avdoshkin:2022xuw,Guo:2022hui,He:2024hkw,He:2025guu} as well as theories defined on time-dependent background spacetimes \cite{Haque:2021hyw,Fan:2022xaa,Adhikari:2022oxr,Bhattacharjee:2022lzy,Li:2024kfm,Chowdhury:2024ntx}. In several notable cases, the Krylov basis coincides with the Fock basis, implying that the Krylov complexity is directly proportional to the average particle number \cite{Adhikari:2022oxr,He:2022ryk,He:2025guu,Haque:2021hyw}. Recent developments have further suggested a deeper connection between Krylov complexity and geometric notions of complexity in holography \cite{Caputa:2021sib,Kar:2021nbm,Rabinovici:2023yex,Zhai:2024tkz}. In particular, it has been proposed that Krylov complexity is proportional to the volume defined via the Fubini–Study metric, lending support to its interpretation as a dual to bulk geometric volume. Motivated by these developments, we investigate particle production arising from quantum fluctuations of a circular cosmic string, adopting the correspondence between Krylov complexity and the average particle number. Within this framework, we investigate whether the resulting Krylov complexity, together with the operator growth rate, exhibits scaling behavior consistent with the “complexity = volume” conjecture. In particular, we aim to determine whether such a correspondence, if present, already emerges at early stages of the evolution or instead becomes manifest only as the string approaches the black hole horizon.

At the classical level, circular string configurations have been extensively studied. For instance, \cite{Larsen:1993mx} analyzed linear perturbations of circular strings in Schwarzschild, Reissner–Nordström, and de Sitter backgrounds, revealing a pronounced radial “ring-collapse” instability as the loop approaches the black hole singularity, while angular perturbations remain bounded. Beyond perturbative dynamics, circular strings have also been investigated in the context of nucleation processes \cite{Garriga:1991tb,Garriga:1992nm} and primordial black hole formation from cosmic strings \cite{Vilenkin:2018zol,Jenkins:2020ctp}. In recent years, observational efforts targeting cosmic strings, particularly through gravitational-wave signatures, have made significant progress \cite{Gouttenoire:2019kij,Sousa:2014gka,Sousa:2016ggw,Blanco-Pillado:2024aca}. Nevertheless, while these classical analyses provide valuable insight into the dynamical evolution and instabilities of circular strings in black hole backgrounds, their quantum aspects, especially quantum fluctuations and particle production near the horizon, remain comparatively unexplored.

Another important motivation for studying particle production near a black hole horizon stems from considerations of wave–particle duality. In Ref.~\cite{Larsen:1998sh}, perturbative analyses indicate that circular strings predominantly spread in the angular direction, while radial spreading is strongly suppressed as the string approaches the horizon due to Lorentz contraction. From the perspective of classical wave dynamics, this suggests that angular modes exhibit a more pronounced wave-like character, whereas radial fluctuations are significantly suppressed. However, invoking wave–particle duality suggests a complementary interpretation: the suppression of radial wave spreading may correspond to an enhancement of particle-like excitations, namely increased particle production originating from radial quantum fluctuations. Conversely, the strong angular spreading may be associated with a reduced particle yield in the angular sector. This observation motivates us to investigate whether radial modes, despite being suppressed at the level of classical wave propagation, can dominate particle production near the horizon, while angular modes contribute less significantly. To our knowledge, a systematic analysis of particle production arising separately from radial and angular fluctuations of an infalling circular string near a black hole horizon has not yet been carried out.

Building on the above analysis and underlying motivations, the organization of this work is as follows. In Section~\ref{StringPerturbaReview}, we revisit the dynamics of a circular cosmic string propagating in a Schwarzschild black hole background, together with the derivation of the quadratic action governing perturbations around the classical string trajectory. In Section~\ref{QuantizationProcedure}, we present a systematic canonical quantization of these quadratic fluctuations within the squeezed-state formalism. In particular, we derive explicit expressions for the squeezing parameters in terms of the mode functions and construct the associated time-evolution operators generated by the quadratic Hermitian Hamiltonian. The corresponding Bogoliubov transformations acting on the creation and annihilation operators are also analyzed. By evolving the initial vacuum state in Fock space, we obtain the resulting two-mode quantum states that characterize the quantum excitations of the string fluctuations. In Section~\ref{ParticleProduction}, we present a detailed construction of the Krylov basis and implement the Lanczos algorithm to compute the Krylov complexity and the associated Lanczos coefficients for quantum fluctuations of an infalling circular string approaching the black hole horizon. We then analyze the time evolution of the Krylov complexity separately for radial and angular perturbation modes. Particular attention is devoted to the near-horizon regime, where the probe string approaches the vicinity of the event horizon, and we investigate the dependence of both the Krylov complexity and the operator growth rate (proportional to the Lanczos coefficients) on the initial size of the probe string, characterized by $GE$. For completeness, the evolution equations for the squeezing parameters, together with technical details of the nested commutator computations, are presented in Appendices~\ref{EvolveSqueezingParameter}–\ref{AppendixNestedCommuta}. In addition, Appendix~\ref{OpeOrderSU11} contains a detailed derivation of the operator-ordering theorems associated with the Lie group $SU(1,1)$. Finally, our conclusions and discussions are presented in Section~\ref{ConcluAndDiscuss}.

\section{Dynamics of a Circular String and the Associated Quadratic Perturbative Action in Schwarzschild Black Hole Spacetime \label{StringPerturbaReview}}

\subsection{Motions of circular string in probe limit}

We begin with the Polyakov action describing a relativistic string,
\begin{align}
\label{BackStringAction}
&S_{(0)}=\frac{-1}{4\pi\alpha^{\prime}}\int d\tau d\sigma\sqrt{-h}h^{\mathtt{A}\mathtt{B}}G_{\mathtt{A}\mathtt{B}}~,
\end{align}where $h_{\mathtt{A}\mathtt{B}}$ denotes the intrinsic world-sheet metric, and $G_{\mathtt{A}\mathtt{B}}$ is the induced metric,
\begin{align}
\label{InduceMetric}
&G_{\mathtt{A}\mathtt{B}}=g_{\mu\nu}\frac{\partial x^{\mu}}{\partial\xi^{\mathtt{A}}}\frac{\partial x^{\nu}}{\xi^{\mathtt{B}}}\,,\,\xi^{\mathtt{A}}=\{\tau,\sigma\}\,,\,x^{\mu}=\{t,r,\theta,\phi\}~,
\end{align}obtained from the spacetime embedding. By varying \eqref{BackStringAction} with respect to the world-sheet metric $h_{\mathtt{A}\mathtt{B}}$ and the embedding coordinates $x^\mu(\xi)$, we obtain the corresponding equations of motion together with the associated constraint equations,
\begin{align}
\label{EOMsBackEmbedding}
0&=h^{\mathtt{A}\mathtt{B}}(\nabla_{\mathtt{A}}\nabla_{\mathtt{B}}x^{\mu}+\Gamma_{\alpha\beta}^{\mu}\frac{\partial x^{\alpha}}{\partial\xi^{\mathtt{A}}}\frac{\partial x^{\beta}}{\partial\xi^{\mathtt{B}}})~,\\
\label{RestrictBackEmbedding}
0&=\frac{1}{2}h_{\mathtt{A}\mathtt{B}}h^{\mathtt{A}_{1}\mathtt{B}_{1}}G_{\mathtt{A}_{1}\mathtt{B}_{1}}-G_{\mathtt{A}\mathtt{B}}~.
\end{align}Without loss of generality, the world-sheet metric may be gauge-fixed to $h_{\mathtt{A}\mathtt{B}}=\eta_{\mathtt{A}\mathtt{B}}$ by exploiting conformal symmetry. To describe a circular string with a time-dependent radius lying in the equatorial plane of a Schwarzschild black hole, we embed the string into spacetime coordinates as
\begin{align}
\label{BackStringEmbed}
&x^{\mu}(\tau,\sigma):t=t(\tau)\,,\,r=r(\tau)\,,\,\theta=\pi/2\,,\,\phi=\sigma~.
\end{align}In this work we are interested in the infall of a circular probe string into a Schwarzschild black hole,
\begin{align}
\label{SchwarzchildBHSol}
ds^{2}&=-f(r)dt^{2}+\frac{dr^{2}}{f(r)}+r^{2}d\theta^{2}+r^{2}\sin^{2}\theta d\phi^{2}~,\\
\nonumber
f(r)&=1-\frac{2\text{G}M}{r}~,
\end{align}under the embedding described by \eqref{BackStringEmbed}. With these setups, the equations of motion \ref{EOMsBackEmbedding} together with the constraints \ref{RestrictBackEmbedding} lead to the following relations,
\begin{align}
\label{eqsStringBackEomT}
0&=\!\ddot{t}(\tau)\!+\frac{f^{\prime}[r(\tau)]}{f[r(\tau)]}\dot{t}(\tau)\dot{r}(\tau)~,\\
\label{eqsStringBackEomR}
0&=\!-\ddot{r}(\tau)\!+\frac{f^{\prime}[r(\tau)]}{2f[r(\tau)]}\dot{r}(\tau)^{2}\!-\frac{f[r(\tau)]}{2}\big(2r(\tau)\!+\!f^{\prime}[r(\tau)]\dot{t}(\tau)^{2}\big)~,\\
\label{RestricStringBack}
0&=\!r(\tau)^{2}\!+\frac{\dot{r}(\tau)^{2}}{f[r(\tau)]}-f[r(\tau)]\,\dot{t}(\tau)^{2}~.
\end{align}After eliminating the $\dot{t}(\tau)^2$ term in \eqref{eqsStringBackEomR} by applying the constraint \eqref{RestricStringBack}, the resulting equation reduces to a linear ODE in the variable $r(\tau)$, whose general solution takes the form
\begin{align}
\label{SolrtauGeneStringBack}
&r(\tau)=\text{G}M+r_{1}\cos(\tau)+r_{2}\sin(\tau)
\end{align}In the present analysis, we are primarily interested in the infalling configurations of a circular string approaching the black hole horizon. Without loss of generality, the initial time may be chosen as $\tau_0=0$. Under this choice, the second term in the solution $r(\tau)$ corresponds to the expanding configurations of the circular string located far from the black hole horizon. Since the term $\sin(\tau)$ falls outside our scope, we discard this branch of the solution by setting the coefficient $r_2=0$ in \eqref{SolrtauGeneStringBack}. On the other hand, by explicitly expanding the Polyakov action in terms of the coordinate embedding \eqref{BackStringEmbed}, we obtain
\begin{align}
S_{(0)}&=\int d\tau\,L[\dot{r},\dot{t},r]=\frac{1}{2\alpha^{\prime}}\int d\tau\,\big\{-r(\tau)^{2}+\frac{\dot{r}(\tau)^{2}}{1-\frac{2\text{G}M}{r(\tau)}}-\big(1-\frac{2\text{G}M}{r(\tau)}\big)\dot{t}(\tau)^{2}\big\}\,.
\end{align}From the Euler–Lagrange equations, it follows directly that the conjugate momentum $\partial L / \partial \dot{t}$ is a conserved quantity. This leads to the relation
\begin{align}
\label{TimeCoorSolBackStringv0}
&\big(1-\frac{2\text{G}M}{r(\tau)}\big)\dot{t}(\tau)=\text{G}E~,
\end{align}where $E$ denotes the energy of the macroscopic string at the initial time $\tau_0$. By substituting \eqref{SolrtauGeneStringBack} and \eqref{TimeCoorSolBackStringv0} back into equations \eqref{eqsStringBackEomT}–\eqref{RestricStringBack}, the previously undetermined coefficient $r_1$ can be determined as $r_{1}^{2}=\text{G}^{2}(M^{2}+E^{2})$. Finally, the time-dependent background trajectory of the string, $\bar{t}(\tau)$ and $\bar{r}(\tau)$, describing the motion of an infalling circular string, can be expressed as
\begin{align}
\label{tbarSol}
&\hspace{-1mm}\bar{t}(\tau)\!=\!\text{G}E\tau\!+\!4\text{G}M\,\text{arctanh}\big(\frac{(M+\sqrt{E^{2}+M^{2}})}{E}\tan(\frac{\tau}{2})\big)~,\\
\label{rbarSol}
&\hspace{-1mm}\bar{r}(\tau)=\text{G}M+\text{G}\sqrt{M^{2}+E^{2}}\cos(\tau)~.
\end{align}Hereafter, we use $\bar{r}(\tau)$ and $\bar{t}(\tau)$ to denote the classical trajectory of the infalling circular string. In the subsequent analysis of string perturbations and quantum fluctuations, all occurrences of the symbols $\bar{r}(\tau)$ and $\bar{t}(\tau)$ refer to the background solutions given in \eqref{tbarSol}–\eqref{rbarSol}.

\subsection{Derivation of the Quadratic Perturbative Action}

In this section, we follow the framework developed in \cite{Larsen:1993mx,Larsen:1998sh,Garriga:1991ts,Guven:1993ex}. From a physical perspective, we are primarily concerned with transverse perturbations, meaning that the deviation $\delta x^{\mu}$ can be expanded in terms of a set of unit vectors orthogonal to the world-sheet of the background string configuration $\bar{x}^\mu (\tau,\sigma)$, namely
\begin{align}
\label{StringPerRestriction}
&\delta x^{\mu}=\bar{n}_{\mathtt{i}}^{\mu}\Phi^{(\mathtt{i})}(\tau,\sigma),\,g_{\mu\nu}\bar{n}_{(\mathtt{i})}^{\mu}\bar{n}_{(\mathtt{j})}^{\nu}\!=\!\delta_{\mathtt{i}\mathtt{j}},\,g_{\mu\nu}\bar{n}_{(\mathtt{i})}^{\mu}\frac{\partial\bar{x}^{\nu}}{\partial\xi^{\mathtt{A}}}\!=\!0~.
\end{align}For a circular string, the set of unit vectors normal to the world-sheet can be decomposed into two independent polarization directions, i.e., the radial ($r$) and polar ($\theta$) components. Accordingly, the polarization indices can be denoted as $\mathtt{i}=r,\theta$, that is,
\begin{align}
&\delta x^{\mu}(\tau,\sigma)=\bar{n}_{(r)}^{\mu}(\tau)\Phi^{(r)}(\tau,\sigma)+\bar{n}_{(\theta)}^{\mu}(\tau)\Phi^{(\theta)}(\tau,\sigma)~.
\end{align}Besides, according to the embedding of the circular string, the quantities $\partial\bar{x}^{\nu}/\partial\xi^{\mathtt{A}}$ can be explicitly expanded as
\begin{align}
\label{VeloTenStringSheet}
&\frac{\partial\bar{x}^{\nu}}{\partial\tau}=(\dot{\bar{t}},\dot{\bar{r}},0,0)~,~\frac{\partial\bar{x}^{\nu}}{\partial\sigma}=(0,0,0,1)~.
\end{align}After substituting \eqref{VeloTenStringSheet} into the constraint conditions \eqref{StringPerRestriction}, we can solve
\begin{small}
\begin{align}
\bar{n}_{(\theta)}^{\mu}(\tau)&=(0,0,\frac{1}{\bar{r}(\tau)},0)~,\\
\nonumber
\bar{n}_{(r)}^{\mu}(\tau)&=\big(-\frac{\dot{\bar{r}}(\tau)}{\bar{r}(\tau)-2\text{G}M},-\frac{\dot{\bar{t}}(\tau)}{\bar{r}(\tau)^2}(\bar{r}(\tau)-2\text{G}M),0,0\big)~,\\
&=\big(\frac{\sqrt{\text{G}^{2}E^{2}+2\text{G}M\bar{r}(\tau)-\bar{r}(\tau)^{2}}}{2\text{G}M-\bar{r}(\tau)},\frac{\text{G}E}{\bar{r}(\tau)},0,0\big)~.
\end{align}
\end{small}In order to derive the quadratic perturbative action more effectively, we need to express the quantities $\frac{\partial\bar{x}^{\nu}}{\partial\tau}$ and $\bar{n}_{(r)}^{\mu}$ in the forms obtained after applying the equations of motion (EOMs) \eqref{eqsStringBackEomT}-\eqref{RestricStringBack}, namely
\begin{align}
&\frac{\partial\bar{x}^{\nu}}{\partial\tau}\overset{\text{EOMs}}{=\!=\!=}(\frac{\text{G}E}{1-2\text{G}M/\bar{r}(\tau)},-\sqrt{\text{G}^{2}E^{2}+2\text{G}M\bar{r}(\tau)-\bar{r}(\tau)^{2}},0,0)~,\\
&\bar{n}_{(r)}^{\mu}(\tau)\overset{\text{EOMs}}{=\!=\!=}\big(\frac{\sqrt{\text{G}^{2}E^{2}+2\text{G}M\bar{r}(\tau)-\bar{r}(\tau)^{2}}}{2\text{G}M-\bar{r}(\tau)},\frac{\text{G}E}{\bar{r}(\tau)},0,0\big)~.
\end{align}So as to reproduce the quadratic-order perturbative action of the circular string, we first need to evaluate the geometric quantities
\begin{align}
&\Omega_{(\mathtt{i})\,\mathtt{A}\mathtt{B}}\!=\!g_{\mu\nu}\bar{n}_{(\mathtt{i})}^{\mu}\frac{\partial\bar{x}^{\alpha}}{\partial\xi^{\mathtt{A}}}(\partial_{\alpha}\frac{\partial\bar{x}^{\nu}}{\partial\xi^{\mathtt{B}}}+\Gamma_{\alpha\lambda}^{\nu}\frac{\partial\bar{x}^{\lambda}}{\partial\xi^{\mathtt{B}}})\big\vert_{\partial_{r}\to\partial_{\bar{r}(\tau)}}^{r\to\bar{r}(\tau),\theta\to\frac{\pi}{2}}~,\\
&\mu_{(\mathtt{i})(\mathtt{j})\,\mathtt{A}}\!=\!g_{\mu\nu}\bar{n}_{(\mathtt{i})}^{\mu}\frac{\partial\bar{x}^{\alpha}}{\partial\xi^{\mathtt{A}}}\big(\partial_{\alpha}\bar{n}_{(\mathtt{j})}^{\nu}+\Gamma_{\alpha\lambda}^{\nu}\bar{n}_{(\mathtt{j})}^{\lambda}\big)\big\vert_{\partial_{r}\to\partial_{\bar{r}(\tau)}}^{r\to\bar{r}(\tau),\theta\to\frac{\pi}{2}}~,
\end{align}where $\Omega_{(\mathtt{i}),\mathtt{A}\mathtt{B}}$ and $\mu_{(\mathtt{i})(\mathtt{j}),\mathtt{A}}$ denote, respectively, the extrinsic curvature tensor of the world sheet and its surface torsion. Based on these geometric constructions, the quadratic action can be written as
\begin{align}
\nonumber
S_{(2)}&=\frac{1}{2\pi\alpha^{\prime}}\int d\tau d\sigma\sqrt{-h}\,\Phi^{(\mathtt{i})}\big(\sum_{\mathtt{k}}h^{\mathtt{A}\mathtt{B}}(\delta_{(\mathtt{i})(\mathtt{k})}\nabla_{\mathtt{A}}+\mu_{(\mathtt{i})(\mathtt{k})~\mathtt{A}})\\&\times(\delta_{(\mathtt{k})(\mathtt{j})}\nabla_{\mathtt{B}}+\mu_{(\mathtt{k})(\mathtt{j})~\mathtt{B}})-\mathcal{V}_{(\mathtt{i})(\mathtt{j})}\big)\Phi^{(\mathtt{j})}~,
\end{align}where $\nabla_{\mathtt{A}}$ represents the covariant derivative associated with the world-sheet metric $h_{\mathtt{A}\mathtt{B}}$, and the potential $\mathcal{V}_{\mathtt{i}\mathtt{j}}$ is defined as
\begin{align}
\label{PolarPotentialQudraticPer}
\mathcal{V}_{(\mathtt{i})(\mathtt{j})}&=h^{\mathtt{A}\mathtt{B}}R_{\mu\alpha\beta\nu}\partial_{\mathtt{A}}\bar{x}^{\mu}\partial_{\mathtt{B}}\bar{x}^{\nu}\bar{n}_{(\mathtt{i})}^{\alpha}\bar{n}_{(\mathtt{j})}^{\beta}\big\vert_{\partial_{r}\to\partial_{\bar{r}(\tau)}}^{r\to\bar{r}(\tau),\theta\to\frac{\pi}{2}}-\frac{2}{G_{~~\mathtt{A}_{1}}^{\mathtt{A}_{1}}}\Omega_{(\mathtt{i})~\mathtt{A}\mathtt{B}}\Omega_{(\mathtt{j})}^{~\mathtt{A}\mathtt{B}}~.
\end{align}Here $G_{~~\mathtt{A}_{1}}^{\mathtt{A}_{1}}$ denotes the trace of the induced metric \eqref{InduceMetric}, while $R_{\mu\alpha\beta\nu}$ is the Riemann curvature tensor associated to the background spacetime \eqref{BackStringEmbed}. To preserve the consistence with the setup in last subsection, we insist on the using of the simplifications $h^{\mathtt{A}\mathtt{B}}\to\eta^{\mathtt{A}\mathtt{B}}$ and $\nabla_{\mathtt{A}}\to\partial_{\mathtt{A}}$. In situation of the flat world-sheet metric $\eta_{\mathtt{A}\mathtt{B}}$ and the Schwarzchild background \eqref{SchwarzchildBHSol}, the non-vanishing components for extrinsic curvature tensor $\Omega_{\mathtt{i}\,\mathtt{A}\mathtt{B}}$ are listed as
\begin{align}
&\Omega_{(r),\tau\tau}=-\text{G}E~,~\Omega_{(r),\sigma\sigma}=-\text{G}E~.
\end{align}Meanwhile, it is straightforward to verify that all components of the surface torsion $\mu_{\mathtt{i}\mathtt{j},\mathtt{A}}$ vanish. Furthermore, upon substituting the above geometric quantities into the potential \eqref{PolarPotentialQudraticPer}, we obtain
\begin{align}
&-\mathcal{V}_{(r)(r)}=\frac{2\text{G}^{2}E^{2}}{\bar{r}(\tau)^{2}}+\frac{\text{G}M}{\bar{r}(\tau)}~,\\
&-\mathcal{V}_{(\theta)(\theta)}=\frac{\text{G}M}{\bar{r}(\tau)}~,\mathcal{V}_{(r)(\theta)}=\mathcal{V}_{(\theta)(r)}=0~.
\end{align}Finally, the resulting quadratic-order effective perturbative action of the string takes the form
\begin{align}
\nonumber
S_{(2)}&=\!\frac{1}{2\pi\alpha^{\prime}}\int\!d\tau d\sigma\big\{\Phi_{(\theta)}\big(\!-\partial_{\tau}^{2}\!+\!\partial_{\sigma}^{2}\!+\!\frac{\text{G}M}{\bar{r}(\tau)}\big)\Phi_{(\theta)}\\
&+\Phi_{(r)}\!\big(\!-\partial_{\tau}^{2}\!+\!\partial_{\sigma}^{2}\!+\!\frac{\text{G}M}{\bar{r}(\tau)}\!+\!\frac{2\text{G}^{2}E^{2}}{\bar{r}(\tau)^{2}}\!\big)\Phi_{(r)}\big\}~,\\
\nonumber
&\equiv\frac{1}{2\pi\alpha^{\prime}}\int d\tau d\sigma\big\{\dot{\Phi}_{(r)}^{2}-\Phi_{(r)}^{\prime2}+\dot{\Phi}_{(\theta)}^{2}-\Phi_{(\theta)}^{\prime2}\\
\label{QuadraticPerString}
&+(\frac{\text{G}M}{\bar{r}(\tau)}+\frac{2\text{G}^{2}E^{2}}{\bar{r}(\tau)^{2}})\Phi_{(r)}^{2}+\frac{\text{G}M}{\bar{r}(\tau)}\Phi_{(\theta)}^{2}\big\}~.
\end{align}

\section{Quantization in squeezed-state formalism \label{QuantizationProcedure}}

In this section, we employ the squeezing formalism to carry out the canonical quantization procedure; further technical details may be found in Ref.\cite{Grain:2019vnq}. Since $\sigma=\phi\in[0,2\pi]$, we can introduce the following Fourier expansions\begin{align}
\label{PerturRadialFourierExpan}
&\Phi_{(r)}(\tau,\sigma)=\frac{1}{2\pi}\sum_{n=+2}^{+\infty}\mathrm{R}_{n}(\tau)\text{e}^{\text{i}n\sigma}+\frac{1}{2\pi}\sum_{n=+2}^{+\infty}\mathrm{R}_{-n}(\tau)\text{e}^{-\text{i}n\sigma}~,~\mathrm{R}_{n}^{\star}(\tau)=\mathrm{R}_{-n}(\tau)~,\\
\label{PerturAngularFourierExpan}
&\Phi_{(\theta)}(\tau,\sigma)=\frac{1}{2\pi}\sum_{n=+2}^{+\infty}\Theta_{n}(\tau)\text{e}^{\text{i}n\sigma}+\frac{1}{2\pi}\sum_{n=+2}^{+\infty}\Theta_{-n}(\tau)\text{e}^{-\text{i}n\sigma}~,~\Theta_{n}^{\star}(\tau)=\Theta_{-n}(\tau)~.
\end{align}The conditions $\mathrm{R}_{n}^{\star}(\Theta_{n}^{\star})=\mathrm{R}_{-n}(\Theta_{-n})$ arise from the requirement that both $\Phi_{(r)}$ and $\Phi_{(\theta)}$ are real scalar fields. In addition, as pointed out in \cite{Garriga:1991ts,Garriga:1991tb,Garriga:1992nm}, the $n=0,1$ modes are excluded, since they correspond merely to rigid translations and rotations that do not alter the shape of the circular string configuration. Consequently, they do not represent physically meaningful oscillations of the string. By exploiting the orthonormality relation $\int_{0}^{2\pi}d\sigma\,\text{e}^{\text{i}(n-m)\sigma}=2\pi\delta_{nm}$, the quadratic-order action \eqref{QuadraticPerString} can be straightforwardly recast in Fourier space upon substituting the mode expansions \eqref{PerturRadialFourierExpan}–\eqref{PerturAngularFourierExpan}. One thus obtains
\begin{align}
\nonumber
S_{(2)}\!=\!\frac{1}{2\alpha^{\prime}\pi^{2}}\sum_{n=+2}^{+\infty}&\int \! d\tau\big\{\vert\dot{\mathrm{R}}_{n}\vert^{2}+(\frac{\text{G}M}{\bar{r}(\tau)}+\frac{2\text{G}^{2}E^{2}}{\bar{r}(\tau)^{2}}-n^{2})\vert\mathrm{R}_{n}\vert^{2}\\
\label{QuadraActionFourier}
&+\vert\dot{\Theta}_{n}\vert^{2}+\big(\frac{\text{G}M}{\bar{r}(\tau)}-n^{2}\big)\vert\Theta_{n}\vert^{2}\big\}~.
\end{align}According to \eqref{QuadraActionFourier}, the corresponding conjugate momentum is constructed as
\begin{align}
\nonumber
&\Pi_{l}^{(\mathrm{R})}(\tau)=\frac{\delta S_{(2)}}{\delta\dot{\mathrm{R}}_{l}^{\star}(\tau)}=\frac{1}{\pi\alpha^{\prime}}\dot{\mathrm{R}}_{l}(\tau)~,\\
\nonumber
&\Pi_{l}^{(\Theta)}(\tau)=\frac{\delta S_{(2)}}{\delta\dot{\Theta}_{l}^{\star}(\tau)}=\frac{1}{\pi\alpha^{\prime}}\dot{\Theta}_{l}(\tau)~,
\end{align}in which we have adopted the following conventions for the functional derivatives
\begin{align}
\nonumber
&\frac{\delta\dot{\mathrm{R}}_{n}^{\star}(\tau^{\prime})}{\delta\dot{\mathrm{R}}_{l}^{\star}(\tau)}=\frac{\delta\dot{\mathrm{R}}_{n}(\tau^{\prime})}{\delta\dot{\mathrm{R}}_{l}(\tau)}=2\pi\delta_{nl}\delta(\tau^{\prime}-\tau)~,\\
\nonumber
&\frac{\delta\dot{\Theta}_{n}^{\star}(\tau^{\prime})}{\delta\dot{\Theta}_{l}^{\star}(\tau)}=\frac{\delta\dot{\Theta}_{n}(\tau^{\prime})}{\delta\dot{\Theta}_{l}(\tau)}=2\pi\delta_{nl}\delta(\tau^{\prime}-\tau)~.
\end{align}And hence the corresponding Hamiltonian is built as
\begin{small}
\begin{align}
\nonumber
H_{(2)}(\tau)\!&=\!\sum_{n=+2}^{+\infty}\big\{\frac{\alpha^{\prime}}{2}\big(\vert\Pi_{n}^{(\mathrm{R})}\vert^{2}\!\!+\!\vert\Pi_{n}^{(\Theta)}\vert^{2}\big)\!+\!\frac{(n^{2}\!\!-\!\text{G}M/\bar{r}(\tau))}{2\pi^{2}\alpha^{\prime}}\vert\mathrm{\Theta}_{n}\vert^{2}\\
\label{QuadraPerHamiltonian}
&+\frac{1}{2\pi^{2}\alpha^{\prime}}(n^{2}-\frac{\text{G}M}{\bar{r}(\tau)}-\frac{2\text{G}^{2}E^{2}}{\bar{r}(\tau)^{2}})\vert\mathrm{R}_{n}\vert^{2}\big\}~.
\end{align}
\end{small}After promoting the fluctuations $\Phi_{(r)},\Pi_{(r)},\Phi_{(\theta)},\Pi_{(\theta)}$ into the quantum field operators, we can build the following mode expansions\begin{small}
\begin{align}
\label{PerturRadialOpeModeExpan}
\widehat{\Phi}_{(r)}(\tau,\sigma)&=\frac{1}{2\pi}\sum_{n=+2}^{n=+\infty}\underbrace{\big(\mathcal{R}_{n}(\tau)\hat{a}_{n}^{(r)}(\tau_{0})+\mathcal{R}_{n}^{\star}(\tau)\hat{a}_{-n}^{(r)\dagger}(\tau_{0})\big)}_{\widehat{\mathrm{R}}_{n}(\tau)}\text{e}^{\text{i}n\sigma}+\frac{1}{2\pi}\sum_{n=+2}^{n=+\infty}\underbrace{\big(\mathcal{R}_{n}(\tau)\hat{a}_{-n}^{(r)}(\tau_{0})+\mathcal{R}_{n}^{\star}(\tau)\hat{a}_{n}^{(r)\dagger}(\tau_{0})\big)}_{\widehat{\mathrm{R}}_{-n}(\tau)=\widehat{\mathrm{R}}_{n}^{\dagger}(\tau)}\text{e}^{-\text{i}n\sigma}~,\\
\label{PerturRadialConMomentumOpeModeExpan}
\hat{\Pi}_{(r)}(\tau,\sigma)&=\frac{1}{2\pi}\sum_{n=+2}^{n=+\infty}\underbrace{\big(\frac{\dot{\mathcal{R}}_{n}(\tau)}{\pi\alpha^{\prime}}\hat{a}_{n}^{(r)}(\tau_{0})+\frac{\dot{\mathcal{R}}_{n}^{\star}(\tau)}{\pi\alpha^{\prime}}\hat{a}_{-n}^{(r)\dagger}(\tau_{0})\big)}_{\widehat{\Pi}_{n}^{(\mathrm{R})}(\tau)}\text{e}^{\text{i}n\sigma}+\frac{1}{2\pi}\sum_{n=+2}^{n=+\infty}\underbrace{\big(\frac{\dot{\mathcal{R}}_{n}^{\star}(\tau)}{\pi\alpha^{\prime}}\hat{a}_{n}^{(r)\dagger}(\tau_{0})+\frac{\dot{\mathcal{R}}_{n}(\tau)}{\pi\alpha^{\prime}}\hat{a}_{-n}^{(r)}(\tau_{0})\big)}_{\widehat{\Pi}_{n}^{(\mathrm{R})\dagger}(\tau)=\widehat{\Pi}_{-n}^{(\mathrm{R})}(\tau)}\text{e}^{-\text{i}n\sigma}~,\\
\label{PerturAngularOpeModeExpan}
\widehat{\Phi}_{(\theta)}(\tau,\sigma)&=\frac{1}{2\pi}\sum_{n=+2}^{n=+\infty}\underbrace{\big(\vartheta_{n}(\tau)\hat{a}_{n}^{(\theta)}(\tau_{0})+\vartheta_{n}^{\star}(\tau)\hat{a}_{-n}^{(\theta)\dagger}(\tau_{0})\big)}_{\hat{\Theta}_{n}(\tau)}\text{e}^{\text{i}n\sigma}+\frac{1}{2\pi}\sum_{n=+2}^{n=+\infty}\underbrace{\big(\vartheta_{n}(\tau)\hat{a}_{-n}^{(\theta)}(\tau_{0})+\vartheta_{n}^{\star}(\tau)\hat{a}_{n}^{(\theta)\dagger}(\tau_{0})\big)}_{\hat{\Theta}_{-n}(\tau)=\hat{\Theta}_{n}^{\dagger}(\tau)}\text{e}^{-\text{i}n\sigma}~,\\
\label{PerturAngularConMomentumOpeModeExpan}
\hat{\Pi}_{(\theta)}(\tau,\sigma)&=\frac{1}{2\pi}\sum_{n=+2}^{n=+\infty}\underbrace{\big(\frac{\dot{\vartheta}_{n}(\tau)}{\pi\alpha^{\prime}}\hat{a}_{n}^{(\theta)}(\tau_{0})+\frac{\dot{\vartheta}_{n}^{\star}(\tau)}{\pi\alpha^{\prime}}\hat{a}_{-n}^{(\theta)\dagger}(\tau_{0})\big)}_{\widehat{\Pi}_{n}^{(\Theta)}(\tau)}\text{e}^{\text{i}n\sigma}+\frac{1}{2\pi}\sum_{n=+2}^{n=+\infty}\underbrace{\big(\frac{\dot{\vartheta}_{n}^{\star}(\tau)}{\pi\alpha^{\prime}}\hat{a}_{n}^{(\theta)\dagger}(\tau_{0})+\frac{\dot{\vartheta}_{n}(\tau)}{\pi\alpha^{\prime}}\hat{a}_{-n}^{(\theta)}(\tau_{0})\big)}_{\widehat{\Pi}_{-n}^{(\Theta)}(\tau)=\widehat{\Pi}_{n}^{(\Theta)\dagger}(\tau)}\text{e}^{-\text{i}n\sigma}~.
\end{align}
\end{small}Note that $\hat{a}_{n}^{(\mathtt{i})}(\tau_0),\hat{a}_{n}^{(\mathtt{i})\dagger}(\tau_0)$ and $\hat{a}_{-n}^{(\mathtt{i})}(\tau_0),\hat{a}_{-n}^{(\mathtt{i})\dagger}(\tau_0)$ constitute two independent sets of creation and annihilation operators, respectively. This implies that the only non-vanishing commutators are only
\begin{small}
\begin{align}
\label{NonVanishCommuTau0}
&2\pi\delta_{nm}\delta^{\mathtt{i}\mathtt{j}}\!=\![\hat{a}_{n}^{(\mathtt{i})}(\tau_{0}),\hat{a}_{m}^{(\mathtt{j})\dagger}(\tau_{0})]\!=\![\hat{a}_{-n}^{(\mathtt{i})}(\tau_{0}),\hat{a}_{-m}^{(\mathtt{j})\dagger}(\tau_{0})]~,
\end{align}
\end{small}where the indices $\mathtt{i},\mathtt{j}$ denote the polarization labels associated with the $r$ and $\theta$ directions. Meanwhile, $\tau_0$ means that they only act on the vacuum state at the initial time $\tau_0$. When handle with the standard canonical quantization procedures, it is required to preserve the equal-time commutator
\begin{align}
\nonumber
\text{i}\delta(\sigma-\sigma^{\prime})&=[\hat{\Phi}_{(r)}(\tau,\sigma),\hat{\Pi}_{(r)}(\tau,\sigma^{\prime})]\\
&=[\hat{\Phi}_{(\theta)}(\tau,\sigma),\hat{\Pi}_{(\theta)}(\tau,\sigma^{\prime})]~.
\end{align}From it, one can derive the Wronskian normalization condition as below
\begin{align}
\nonumber
\text{i}\pi\alpha^{\prime}&=\mathcal{R}_{n}(\tau)\dot{\mathcal{R}}_{n}^{\star}(\tau)-\mathcal{R}_{n}^{\star}(\tau)\dot{\mathcal{R}}_{n}(\tau)\\
\label{WronskianMode}
&=\vartheta_{n}(\tau)\dot{\vartheta}_{n}^{\star}(\tau)-\vartheta_{n}^{\star}(\tau)\dot{\vartheta}_{n}(\tau)\, , \, n\geq 2~.
\end{align}Furthermore, the time evolution of the mode functions is governed by the following equation of motion:
\begin{align}
\label{PerLinearModeR}
&-\ddot{\mathcal{R}}_n+(\frac{\text{G}M}{\bar{r}(\tau)}+\frac{2\text{G}^{2}E^{2}}{\bar{r}(\tau)^{2}}-n^{2})\mathcal{R}_n=0~,\\
\label{NumSolAngularMode}
&-\ddot{\vartheta}_{n}+(\frac{\text{G}M}{\bar{r}(\tau)}-n^{2})\mathrm{\vartheta}_{n}=0~.
\end{align}In addition, by examining the commutators of the field–operator variables, we note that they indeed assemble into the standard symplectic structure
\begin{small}
\begin{align}
\label{SympleFourierRadial}
&\hspace{-1mm}\left(\begin{array}{cc}
[\widehat{\mathrm{R}}_{n}(\tau),\widehat{\mathrm{R}}_{l}^{\dagger}(\tau)] & [\widehat{\mathrm{R}}_{n}(\tau),\widehat{\Pi}_{l}^{(\mathrm{R})\dagger}(\tau)]\\{}
[\widehat{\Pi}_{n}^{(\mathrm{R})}(\tau),\widehat{\mathrm{R}}_{l}^{\dagger}(\tau)] & [\widehat{\Pi}_{n}^{(\mathrm{R})}(\tau),\widehat{\Pi}_{l}^{(\mathrm{R})\dagger}(\tau)]
\end{array}\right)\!\!=\!2\pi\text{i}\delta_{nl}\boldsymbol{\Omega}_{2\times2}~,\\
\nonumber
&\quad\quad\quad\quad\quad\quad\quad\quad\boldsymbol{\Omega}_{2\times2}=\left(\begin{array}{cc}
0 & 1\\
-1 & 0
\end{array}\right)~.
\end{align}
\end{small}It is important to emphasize that the symplectic structure derived in the quantization procedure for the radial fluctuation $\Phi_{(r)}$ applies equally to the angular fluctuation $\Phi_{(\theta)}$. Finally, we emphasize that, in principle, the four–dimensional phase–space vector
\begin{align}
&\hat{\vec{z}}_{n}(\tau)=\big(\hat{\mathrm{R}}_{n}(\tau),\hat{\Theta}_{n}(\tau),\hat{\Pi}_{n}^{(\mathrm{R})}(\tau),\hat{\Pi}_{n}^{(\Theta)}(\tau)\big)^{T}~,
\end{align}forms an $Sp(4,\mathbb{R})$ symplectic structure. However, because the quadratic perturbative action \eqref{QuadraticPerString} shows that the radial fluctuation $\Phi_{(r)}$ and the angular fluctuation $\Phi_{(\theta)}$ are completely decoupled from each other, the actual symmetry of the system reduces to
\begin{align}
&Sp(2,R)\oplus Sp(2,R)\in Sp(4,R)~,
\end{align}in which the symmetry group $Sp(2,\mathbb{R})\oplus Sp(2,\mathbb{R})$ possesses 6 degrees of freedom, whereas the full group $Sp(4,\mathbb{R})$ contains 10 degrees of freedom. In other words, we only need to consider two independent two-dimensional phase–space vectors, namely
\begin{align}
&\big(\hat{\mathrm{R}}_{n}(\tau),\hat{\Pi}_{n}^{(\mathrm{R})}(\tau)\big)^{T}~,~\big(\hat{\Theta}_{n}(\tau),\hat{\Pi}_{n}^{(\Theta)}(\tau)\big)^{T}~,
\end{align}separately. This is precisely why, throughout the quantization procedure above, the radial and angular perturbations can be treated as two mutually independent two-dimensional phase–space sectors.

\subsection{Dynamics of Creation and Annihilation Operators in the Heisenberg Picture and the two-mode quantum state Formalism}

\begin{figure}[H]
 	\begin{center}
 		\includegraphics[scale=0.345]{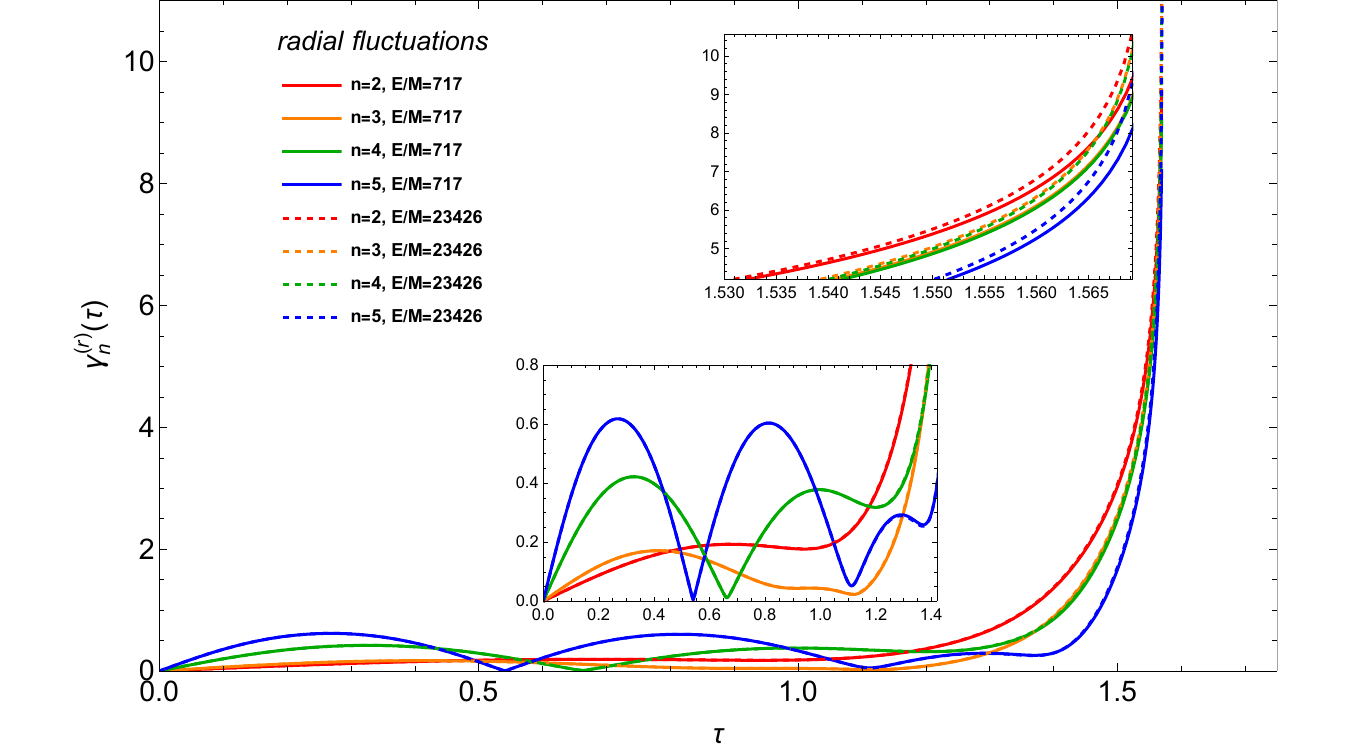}
        \includegraphics[scale=0.345]{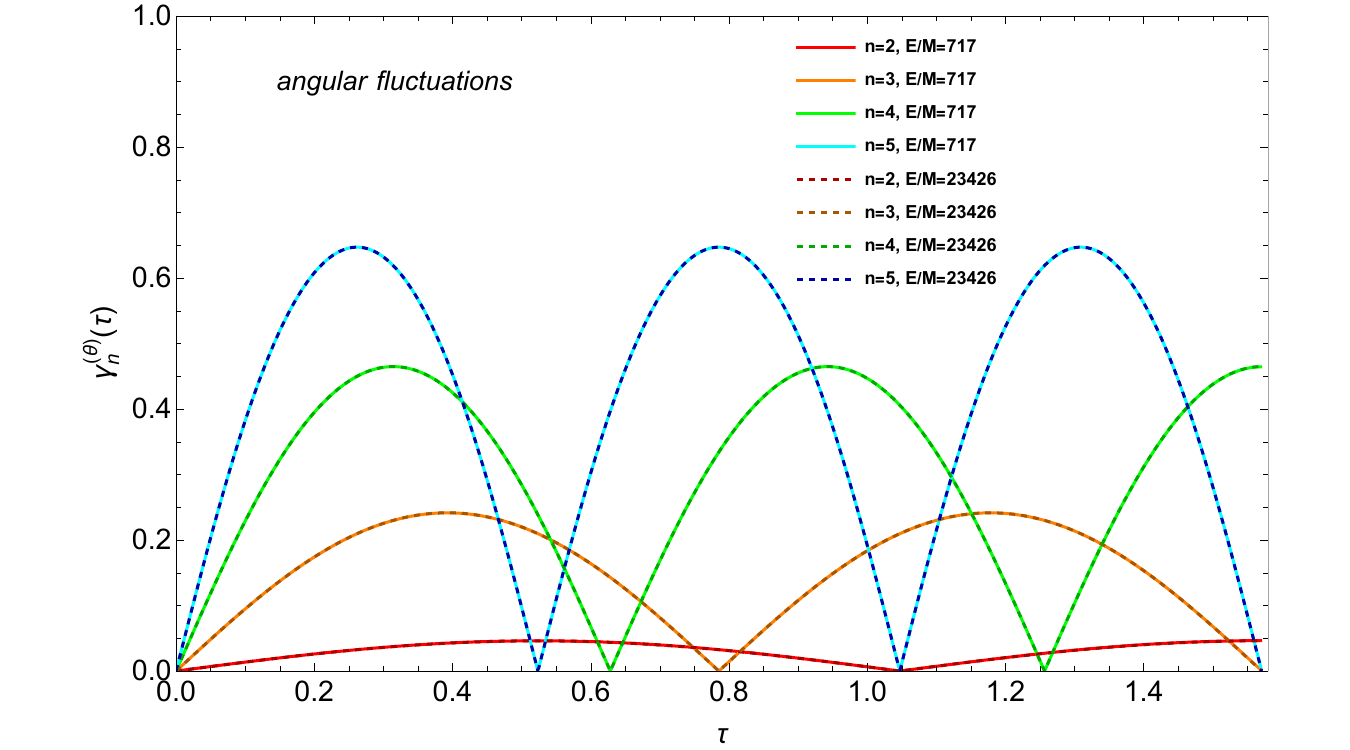}
 		\caption{Time evolution of the squeezing amplitude $\gamma^{(\mathtt{i})}_n(\tau)$ for several winding numbers $n$ in the regime $E/M \gg 1$, shown for two representative values of $E$. In the numerical analysis, the parameters are fixed to $\text{G}=10^4$ and $M=1$.}
 		\label{SqueezAmplitudeDifferentE}
 	\end{center}
 \end{figure}

After performing a linear canonical transformation, it is convenient to work in the helicity basis \cite{Grain:2019vnq}, in which a pair of time-dependent creation and annihilation operators can be constructed as
\begin{align}
\label{AssumeRadialAnnihilaTauFinal}
&\hat{a}_{\pm n}^{(r)}(\tau)=\frac{1}{\sqrt{2\alpha^{\prime}}}\hat{\mathrm{R}}_{\pm n}(\tau)+\text{i}\sqrt{\frac{\alpha^{\prime}}{2}}\hat{\Pi}_{\pm n}^{(\mathrm{R})}(\tau)~,\\
\label{AssumeRadialCreationTauFinal}
&\hat{a}_{\mp n}^{(r)\dagger}(\tau)=\frac{1}{\sqrt{2\alpha^{\prime}}}\hat{\mathrm{R}}_{\mp n}^{\dagger}(\tau)-\text{i}\sqrt{\frac{\alpha^{\prime}}{2}}\hat{\Pi}_{\mp n}^{(\mathrm{R})\dagger}(\tau)~,\\
\label{AssumeAngularAnnihilaTauFinal}
&\hat{a}_{\pm n}^{(\theta)}(\tau)=\frac{1}{\sqrt{2\alpha^{\prime}}}\hat{\Theta}_{\pm n}(\tau)+\text{i}\sqrt{\frac{\alpha^{\prime}}{2}}\hat{\Pi}_{\pm n}^{(\Theta)}(\tau)~,\\
\label{AssumeAngularCreationTauFinal}
&\hat{a}_{\mp n}^{(\theta)\dagger}(\tau)=\frac{1}{\sqrt{2\alpha^{\prime}}}\hat{\Theta}_{\mp n}^{\dagger}(\tau)-\text{i}\sqrt{\frac{\alpha^{\prime}}{2}}\hat{\Pi}_{\mp n}^{(\Theta)\dagger}(\tau)~.
\end{align}For these time-dependent operators, the commutation relations in \eqref{NonVanishCommuTau0} can be consistently generalized to an arbitrary time slice, yielding
\begin{align}
\label{TimeDependCreaAnnihilaCommu}
&2\pi\delta_{nm}\delta^{\mathtt{i}\mathtt{j}}\!=\![\hat{a}_{n}^{(\mathtt{i})}(\tau),\hat{a}_{m}^{(\mathtt{j})\dagger}(\tau)]\!=\![\hat{a}_{-n}^{(\mathtt{i})}(\tau),\hat{a}_{-m}^{(\mathtt{j})\dagger}(\tau)]~,
\end{align}while all other commutators vanish identically. It is also straightforward to verify that the operators $\hat{a}_{\pm n}^{(\mathtt{i})}(\tau)$ and $\hat{a}_{\pm n}^{(\mathtt{j})\dagger}(\tau)$ are dimensionless, as follows from the dimensional analysis
\begin{align}
\nonumber
&\text{Dim}[\tau]=\text{Dim}[\sigma]=\text{Dim}[\hat{a}_{\pm n}^{(\mathtt{i})}(\tau)]=\text{Dim}[\hat{a}_{\pm n}^{(\mathtt{i})\dagger}(\tau)]=\Lambda^0~,\\
\nonumber
&\text{Dim}[\frac{1}{\alpha^{\prime}}]=\Lambda^{-2}~,~\text{Dim}[\hat{\mathrm{R}}_{n}(\tau)]=\text{Dim}[\hat{\Theta}_{n}(\tau)]=\Lambda~,\\
\nonumber
&\text{Dim}[\hat{\Pi}_{n}^{(\mathrm{R})}(\tau)]=\text{Dim}[\hat{\Pi}_{n}^{(\Theta)}(\tau)]=\Lambda^{-1}~.
\end{align}Within this framework, the initial conditions for the mode functions $\mathcal{R}{n}(\tau)$ and $\vartheta{n}(\tau)$ are fixed by the operator definitions \eqref{AssumeRadialAnnihilaTauFinal}–\eqref{AssumeAngularCreationTauFinal}, namely
\begin{align}
\label{InitialConditionModes}
&\vartheta_{n}(\tau_{0})\!=\!\frac{\text{i}}{\pi}\dot{\vartheta}_{n}(\tau_{0})\!=\!\mathcal{R}_{n}(\tau_{0})\!=\!\frac{\text{i}}{\pi}\dot{\mathcal{R}}_{n}(\tau_{0})\!=\!\sqrt{\frac{\alpha^{\prime}}{2}}~.
\end{align}It is important to note that this initial condition is consistent with the Wronskian normalization \eqref{WronskianMode}. The combination of this normalization condition and the chosen initial condition is not arbitrary; rather, it ensures a well-defined vacuum state at the initial time $\tau_0$. As will be shown later, this choice indeed guarantees that the mean particle number vanishes at $\tau_0$. Furthermore, by substituting these time-dependent operators back into \eqref{QuadraPerHamiltonian}, the Hamiltonian operator can be expressed in terms of the time-dependent creation and annihilation operators as\begin{small}
\begin{align}
\nonumber
\hat{\mathcal{H}}^{\text{(2)}}\!=\!\hat{\mathcal{H}}_{(r)}^{\text{(2)}}+\hat{\mathcal{H}}_{(\theta)}^{\text{(2)}}&=\!\frac{1}{4\pi^{2}}\sum_{n=+2}^{+\infty}\bigg\{(n^{2}\!-\!\frac{\text{G}M}{\bar{r}}-\frac{2\text{G}^{2}E^{2}}{\bar{r}^{2}}\!+\!\pi^{2})\big( \hat{a}_{n}^{(r)\dagger}(\tau)\hat{a}_{n}^{(r)}(\tau)\!+\!\hat{a}_{-n}^{(r)}(\tau)\hat{a}_{-n}^{(r)\dagger}(\tau) \big)\\
\nonumber
&+(n^{2}\!-\!\frac{\text{G}M}{\bar{r}}\!-\!\frac{2\text{G}^{2}E^{2}}{\bar{r}^{2}}-\pi^{2})\big(\hat{a}_{n}^{(r)\dagger}(\tau)\hat{a}_{-n}^{(r)\dagger}(\tau)\!+\!\hat{a}_{-n}^{(r)}(\tau)\hat{a}_{n}^{(r)}(\tau) \big)\\
\nonumber
&+(n^{2}-\frac{\text{G}M}{\bar{r}}+\pi^{2})\big(\hat{a}_{n}^{(\theta)\dagger}(\tau)\hat{a}_{n}^{(\theta)}(\tau)+\hat{a}_{-n}^{(\theta)}(\tau)\hat{a}_{-n}^{(\theta)\dagger}(\tau)\big)\\
\label{HamilQuadraticPer}
&+(n^{2}-\frac{\text{G}M}{\bar{r}}-\pi^{2})(\hat{a}_{n}^{(\theta)\dagger}(\tau)\hat{a}_{-n}^{(\theta)\dagger}(\tau)+\hat{a}_{-n}^{(\theta)}(\tau)\hat{a}_{n}^{(\theta)}(\tau)\big)\bigg\}~.
\end{align}
\end{small}Note that the $SU(1,1)$ Lie group symmetry structure of the Hamiltonian operator \eqref{HamilQuadraticPer} is analyzed in detail in Appendix~\ref{EvolveSqueezingParameter}, in particular in the derivations presented in \eqref{SU11LinearMatrxGenerator}–\eqref{LieAlgebraFormalismQuadraHamil}. In principle, once the quadratic Hamiltonian \eqref{HamilQuadraticPer} is specified, the time-evolution operator $\hat{\mathcal{U}}(\tau,\tau_{0})$ can be formally written as
\begin{align}
\label{DefineTimeEvoluOperator}
&\hat{\mathcal{U}}^{(\mathtt{i})}(\tau,\tau_{0})=\mathcal{T}\exp\big(-\text{i}\int_{\tau_{0}}^{\tau}d\tilde{\tau}\,\hat{\mathcal{H}}_{(\mathtt{i})}^{\text{(2)}}(\tilde{\tau})\big)~.
\end{align}In practice, however, an explicit evaluation of this operator by expanding the exponential in terms of creation and annihilation operators is highly nontrivial. Such an approach typically involves the Zassenhaus decomposition and the computation of a hierarchy of nested commutators, rendering it technically unwieldy. A more efficient strategy is to exploit the underlying algebraic structure: the $su(1,1)$ Lie algebra offers a natural and systematic framework for constructing the time-evolution operator in a tractable manner. According to the left-polar decomposition of a $SU(1,1)$ group element \cite{Barnett,Puri:2001}, it yields
\begin{align}
\label{TimeEvolutionOpeDecompose}
&\hat{\mathcal{U}}^{(\mathtt{i})}(\tau,\tau_{0})\!=\!\underbrace{\exp\bigg(\frac{1}{2\pi}\sum_{n=2}^{+\infty}\big(\xi_{n}^{(\mathtt{i})}(\tau)\hat{\mathcal{K}}_{+,n}^{(\mathtt{i})}(\tau_{0})-\xi_{n}^{(\mathtt{i})\star}(\tau)\hat{\mathcal{K}}_{-,n}^{(\mathtt{i})}(\tau_{0})\big)\bigg)}_{\hat{\mathcal{S}}^{(\mathtt{i})}}\underbrace{\exp\big(\frac{\text{i}}{\pi}\sum_{n=2}^{+\infty}\varpi_{n}^{(\mathtt{i})}(\tau)\hat{\mathcal{K}}_{z,n}^{(\mathtt{i})}(\tau_{0})\big)}_{\hat{\mathcal{R}}^{(\mathtt{i})}}~,
\end{align}in which the time-dependent parameter $\xi_{n}^{(\mathtt{i})}$ is conveniently parametrized as $\xi_{m}^{(\mathtt{i})}(\tau)=-\text{i}\gamma_{m}^{(\mathtt{i})}(\tau)\text{e}^{2\text{i}\varphi_{m}^{(\mathtt{i})}(\tau)}$. Within the squeezed-state formalism, $\gamma_{m}^{(\mathtt{i})}(\tau)$ and $\varphi_{m}^{(\mathtt{i})}(\tau)$ correspond to the squeezing amplitude and squeezing phase, respectively, while $\varpi_{m}^{(\mathtt{i})}(\tau)$ denotes the rotation angle. Accordingly, the operators $\hat{\mathcal{S}}\big(\gamma(\tau),\varphi(\tau)\big)$ and $\hat{\mathcal{R}}\big(\varpi(\tau)\big)$ represent the two-mode squeezing operator and the rotation operator. Furthermore, by invoking standard operator-ordering theorems \cite{Barnett,Puri:2001}, the squeezing operator can be factorized into a disentangled form as
\begin{small}
\begin{align}
\nonumber
&\hat{\mathcal{S}}^{(\mathtt{i})}\big(\gamma,\varphi\big)\!=\!\exp\big\{\frac{1}{2\pi}\sum_{m=2}^{+\infty}\big(\text{e}^{2\text{i}\varphi_{m}^{(\mathtt{i})}}\tanh(\gamma_{m}^{(\mathtt{i})})\hat{a}_{m}^{(\mathtt{i})\dagger}(\tau_{0})\hat{a}_{-m}^{(\mathtt{i})\dagger}(\tau_{0})\big)\big\}\\
\nonumber
&\cdot\exp\big\{\frac{1}{2\pi}\sum_{m=2}^{+\infty}\big(-\ln(\cosh(\gamma_{m}^{(\mathtt{i})}))\big(\hat{a}_{-m}^{(\mathtt{i})}\hat{a}_{-m}^{(\mathtt{i})\dagger}+\hat{a}_{m}^{(\mathtt{i})\dagger}\hat{a}_{m}^{(\mathtt{i})}\big)\big)\big\}\\
&\cdot\exp\big\{\frac{1}{2\pi}\sum_{m=2}^{+\infty}\big(\!-\!\text{e}^{-2\text{i}\varphi_{m}^{(\mathtt{i})}}\tanh(\gamma_{m}^{(\mathtt{i})})\hat{a}_{-m}^{(\mathtt{i})}(\tau_{0})\hat{a}_{m}^{(\mathtt{i})}(\tau_{0})\big)\big\}~,
\end{align}
\end{small}For a detailed derivation of the operator-ordering theorems associated with the Lie group $SU(1,1)$, we refer the reader to Refs.~\cite{Barnett,Puri:2001,Grain:2019vnq,Martin:2015qta,Li:2021kfq,Liu:2021nzx}. For completeness, we also provide explicit derivations in Appendix~\ref{OpeOrderSU11}. On the other hand, in addition to the linear canonical transformations~\eqref{AssumeRadialAnnihilaTauFinal}-\eqref{AssumeAngularCreationTauFinal}, the time evolution of the creation and annihilation operators can be constructed with the aid of the evolution operator $\hat{\mathcal{U}}^{(\mathtt{i})}(\tau,\tau_{0})$. In particular, the operators at an arbitrary time $\tau$, namely $\hat{a}_{\pm n}^{(\mathtt{i})}(\tau)$ and $\hat{a}_{\mp n}^{(\mathtt{i})\dagger}(\tau)$, are given by \eqref{ConstruAnnihilaOpeAnyTau}-\eqref{ConstruCreationOpeAnyTau}. By repeatedly applying the commutator expansion formula
$\text{e}^{\hat{B}}\hat{A}\text{e}^{-\hat{B}}=\hat{A}+[\hat{B},\hat{A}]+\frac{1}{2!}[\hat{B},[\hat{B},\hat{A}]]+\dots$,
as detailed in Appendix~\ref{AppendixNestedCommuta}, one obtains\begin{align}
\label{CreationOpeAnyTauSqueeze}
&\hat{a}_{n}^{(\mathtt{i})}(\tau)=\cosh(\gamma_{n}^{(\mathtt{i})}(\tau))\text{e}^{\text{i}\varpi_{n}^{(\mathtt{i})}(\tau)}\hat{a}_{n}^{(\mathtt{i})}(\tau_{0})+\text{e}^{-\text{i}\varpi_{n}^{(\mathtt{i})}(\tau)}\text{e}^{2\text{i}\varphi_{n}^{(\mathtt{i})}(\tau)}\sinh(\gamma_{n}^{(\mathtt{i})}(\tau))\hat{a}_{-n}^{(\mathtt{i})\dagger}(\tau_{0}),\\
\label{AnnihiOpeAnyTauSqueeze}
&\hat{a}_{-n}^{(\mathtt{i})\dagger}(\tau)=\cosh(\gamma_{n}^{(\mathtt{i})}(\tau))\text{e}^{-\text{i}\varpi_{n}^{(\mathtt{i})}(\tau)}\hat{a}_{-n}^{(\mathtt{i})\dagger}(\tau_{0})+\text{e}^{\text{i}\varpi_{n}^{(\mathtt{i})}(\tau)}\text{e}^{-2\text{i}\varphi_{n}^{(\mathtt{i})}(\tau)}\sinh(\gamma_{n}^{(\mathtt{i})}(\tau))\hat{a}_{n}^{(\mathtt{i})}(\tau_{0})~.
\end{align}By matching the expressions in \eqref{CreationOpeAnyTauSqueeze}--\eqref{AnnihiOpeAnyTauSqueeze} with those derived from the linear canonical transformations~\eqref{AssumeRadialAnnihilaTauFinal}--\eqref{AssumeRadialCreationTauFinal}, we obtain the relations between the squeezing parameters and the mode functions,
\begin{small}
\begin{align}
\label{SqueeAngleToModeFunc}
&\cos(2\varphi_{n}^{(r)})\!=\!\frac{\text{Re}\big\{\!\big(\mathcal{R}_{n}(\tau)\!+\!\frac{\text{i}}{\pi}\!\dot{\mathcal{R}}_{n}(\tau)\big)\big(\mathcal{R}_{n}^{\star}(\tau)\!+\!\frac{\text{i}}{\pi}\!\dot{\mathcal{R}}_{n}^{\star}(\tau)\big)\!\big\}}{\big\vert\mathcal{R}_{n}(\tau)\!+\!\frac{\text{i}}{\pi}\!\dot{\mathcal{R}}_{n}(\tau)\vert\cdot\vert\mathcal{R}_{n}(\tau)\!-\!\frac{\text{i}}{\pi}\!\dot{\mathcal{R}}_{n}(\tau)\vert}~,\\
\label{SqueeAmpliToModeFunc}
&\sinh^{2}(\gamma_{n}^{(r)})=\frac{1}{2\alpha^{\prime}}\vert\mathcal{R}_{n}(\tau)-\frac{\text{i}}{\pi}\dot{\mathcal{R}}_{n}(\tau)\vert^{2}~,\\
\label{RotationFactorToModeFunc}
&\cos(\varpi_{n}^{(r)})=\frac{\text{Re}\big(\mathcal{R}_{n}(\tau)+\frac{\text{i}}{\pi}\dot{\mathcal{R}}_{n}(\tau)\big)}{\big\vert\mathcal{R}_{n}(\tau)+\frac{\text{i}}{\pi}\dot{\mathcal{R}}_{n}(\tau)\vert}~.
\end{align}
\end{small}An identical set of expressions is obtained for fluctuations in the angular polarization sector upon the replacement $\mathcal{R}_n \to \vartheta_{n}$. Given the initial conditions in \eqref{InitialConditionModes}, it follows directly that $\gamma_{n}^{(\mathtt{i})}(\tau_{0})=\varpi_{n}^{(\mathtt{i})}(\tau_{0})=\varphi_{n}^{(\mathtt{i})}(\tau_{0})=0$. After numerically solving the equations of motion~\eqref{PerLinearModeR}--\eqref{NumSolAngularMode} and substituting the solutions into the squeezing parameters, in particular the amplitude $\gamma_{n}^{(\mathtt{i})}(\tau)$ as defined in~\eqref{SqueeAmpliToModeFunc}, we obtain the results shown in Fig.~\ref{SqueezAmplitudeDifferentE}. The accuracy of the numerical solutions is monitored using the normalization conditions in \eqref{WronskianMode}. In Fig.~\ref{SqueezAmplitudeDifferentE}, we compare the time evolution of the quantum-state amplitudes $\gamma^{(\mathtt{i})}_n(\tau)$ for representative winding numbers at different values of $E$, while keeping the Schwarzschild black hole mass $M$ fixed. The initial size of the infalling string is given by $G(M+\sqrt{M^2+E^2})$, implying that larger values of $E$ correspond to strings with a larger initial spatial extent. We find that the amplitudes associated with angular fluctuations remain parametrically small, even as the circular string approaches the event horizon, and exhibit negligible dependence on $E$. In contrast, the amplitudes for radial fluctuations, $\gamma_n^{(r)}(\tau)$, undergo a rapid growth as the string approaches the black hole horizon. From a physical perspective, this behavior can be interpreted as an effective thermalization of the circular string in the near-horizon region. The growth of $\gamma_n^{(r)}(\tau)$ reflects the transition of the underlying quantum state from a two-mode squeezed vacuum toward a thermofield double state. Notably, the dependence of $\gamma_n^{(r)}(\tau)$ on $E$ becomes appreciable only after the system enters this thermally dominated regime.
In addition, another physical quantity that will play an important role in the subsequent analysis is the particle number. We therefore derive its explicit expression here,
 \begin{align}
\mathcal{N}_{n}(\tau)&=\frac{1}{2\pi}\sum_{m=2}^{+\infty}\langle\Psi_{\gamma,\varphi,\varpi}(\tau)\vert\hat{a}_{n}^{\dagger}(\tau_{0})\hat{a}_{m}(\tau_{0})\vert\Psi_{\gamma,\varphi,\varpi}(\tau)\rangle\\
\nonumber
&=\frac{1}{2\pi}\sum_{m=2}^{+\infty}\langle\tilde{0}_{n},\tilde{0}_{-n}\vert\hat{\mathcal{U}}^{\dagger}(\tau;\tau_{0})\hat{a}_{n}^{\dagger}(\tau_{0})\hat{\mathcal{U}}(\tau;\tau_{0})\cdot\hat{\mathcal{U}}^{\dagger}(\tau;\tau_{0})\hat{a}_{m}(\tau_{0})\hat{\mathcal{U}}(\tau;\tau_{0})\vert\tilde{0}_{m},\tilde{0}_{-m}\rangle_{\tau_{0}}~.
\end{align}In Appendix~\ref{AppendixNestedCommuta}, we present a detailed derivation of how the time-evolution operator acts on $\hat{a}_{\pm n}(\tau_0)$ and $\hat{a}^\dagger_{\pm n}(\tau_0)$. By employing the results obtained in \eqref{TimeEvolutionAnnihi}--\eqref{TimeEvolutionCreation}, we then obtain
\begin{align}
\nonumber
\mathcal{N}_{n}(\tau)&=\frac{1}{2\pi}\sum_{m=2}^{+\infty}\sinh(\gamma_{n}(\tau))\sinh(\gamma_{m}(\tau))\text{e}^{\text{i}(\varpi_{n}(\tau)-\varpi_{m}(\tau)-2\varphi_{n}(\tau)+2\varphi_{m}(\tau))}\\
\label{DefineParticleNum}
&\times\langle\tilde{0}_{n},\tilde{0}_{-n}\vert\hat{a}_{-n}(\tau_{0})\hat{a}_{-m}^{\dagger}(\tau_{0})\vert\tilde{0}_{m},\tilde{0}_{-m}\rangle_{\tau_{0}}=\sinh^{2}(\gamma_{n}(\tau))~.
\end{align}Using the expression for the particle number, $\mathcal{N}_{n}^{(\mathtt{i})}(\tau)=\sinh(\gamma_{n}^{(\mathtt{i})}(\tau))^{2}$, together with the time evolution of $\gamma_{n}^{(\mathtt{i})}(\tau)$ shown in Fig.~\ref{SqueezAmplitudeDifferentE}, one finds that, as the circular string approaches the event horizon, a significant number of particles are produced in the radial fluctuation modes, whereas particle production in the angular modes remains strongly suppressed. From the perspective of wave–particle duality, this behavior admits a complementary interpretation consistent with the results reported in \cite{Larsen:1998sh}. In particular, near the event horizon, radial quantum fluctuations exhibit an increasingly particle-like character, leading to enhanced particle production accompanied by a suppression of the spatial spreading of the corresponding wave packets. By contrast, angular fluctuations display the opposite trend: particle production remains inefficient, while their wave-like nature becomes more pronounced.

As derived in \eqref{GeneDefineSqueezed}–\eqref{TwoModeQuantumStateAnyTau}, the time-evolution operator $\hat{\mathcal{U}}^{(\mathtt{i})}(\tau,\tau_{0})$ generates the two-mode quantum state associated with a given winding number $n$, which can be written as
\begin{align}
\nonumber
\vert\Psi_{\gamma_{n},\varphi_{n},\varpi_{n}}^{(\mathtt{i})}(\tau)\rangle&=\hat{\mathcal{U}}^{(\mathtt{i})}(\tau;\tau_{0})\vert\tilde{0}_{n},\tilde{0}_{-n}\rangle_{\tau_{0}}=\hat{\mathcal{S}}^{(\mathtt{i})}(\gamma,\varphi)\hat{\mathcal{R}}^{(\mathtt{i})}(\varpi)\vert\tilde{0}_{n},\tilde{0}_{-n}\rangle_{\tau_{0}}\\
\label{TwoModeQuantumStateAnyTau}
&=\frac{\text{e}^{\text{i}\varpi_{n}^{(\mathtt{i})}(\tau)}}{\cosh\big(\gamma_{n}^{(\mathtt{i})}(\tau)\big)}\sum_{\tilde{m}=0}^{\infty}\bigg(\text{e}^{2\text{i}\varphi_{n}^{(\mathtt{i})}(\tau)}\tanh\big(\gamma_{n}^{(\mathtt{i})}(\tau)\big)\bigg)^{\tilde{m}}\big\vert\tilde{m}_{n},\tilde{m}_{-n}\rangle_{\tau_{0}}~.
\end{align}
Here, $\tilde{m}$ and $\tilde{n}$ denote the occupation numbers of the corresponding Fock states. This notation is adopted to clearly distinguish them from the winding numbers $m$ and $n$ that label the Fourier modes in \eqref{PerturRadialFourierExpan}–\eqref{PerturAngularFourierExpan}. In fact, as the string approaches arbitrarily close to the event horizon, the two-mode quantum state \eqref{TwoModeQuantumStateAnyTau} transforms, up to an overall phase, into a thermofield double state \cite{Haque:2021kdm},
\begin{align}
&\vert\text{TFD}\rangle_{\varphi^{\star},\varpi^{\star}}=\prod_{m=1}^{\infty}\frac{1}{\sqrt{\mathcal{Z}_{m}}}\sum_{\tilde{n}=0}^{\infty}\big(\text{e}^{\text{i}(2\tilde{n}\varphi_{m}^{\star}+\varpi_{m}^{\star})}\text{e}^{-\tilde{n}\beta\omega_{m}/2}\big)\vert\tilde{n}_{m},\tilde{n}_{-m}\rangle~,
\end{align}where $\omega_{m}$ denotes the frequency of particles excited by string fluctuations with winding number $m$, and $\beta$ is the inverse Hawking temperature of the black hole, i.e. $\beta = \frac{1}{T} = 8\pi \text{G} M$. Consequently, at $\tau_{h}$ the squeezing amplitude takes the form
\begin{align}
&\gamma_{m}(\tau_{h})=\text{arctanh}\big(\exp(-\beta\omega_{m}/2)\big)~,
\end{align}and the corresponding particle number evaluates to
\begin{align}
&\mathcal{N}_{m}(\tau_{h})=\sinh^{2}\big(\gamma_{m}(\tau_{h})\big)=\frac{1}{\exp(\beta\omega_{m})-1}~,
\end{align}which coincides with the thermal Bose–Einstein distribution characteristic of Hawking radiation.

\begin{figure}[H]
 	\begin{center}
 		\includegraphics[scale=0.5]{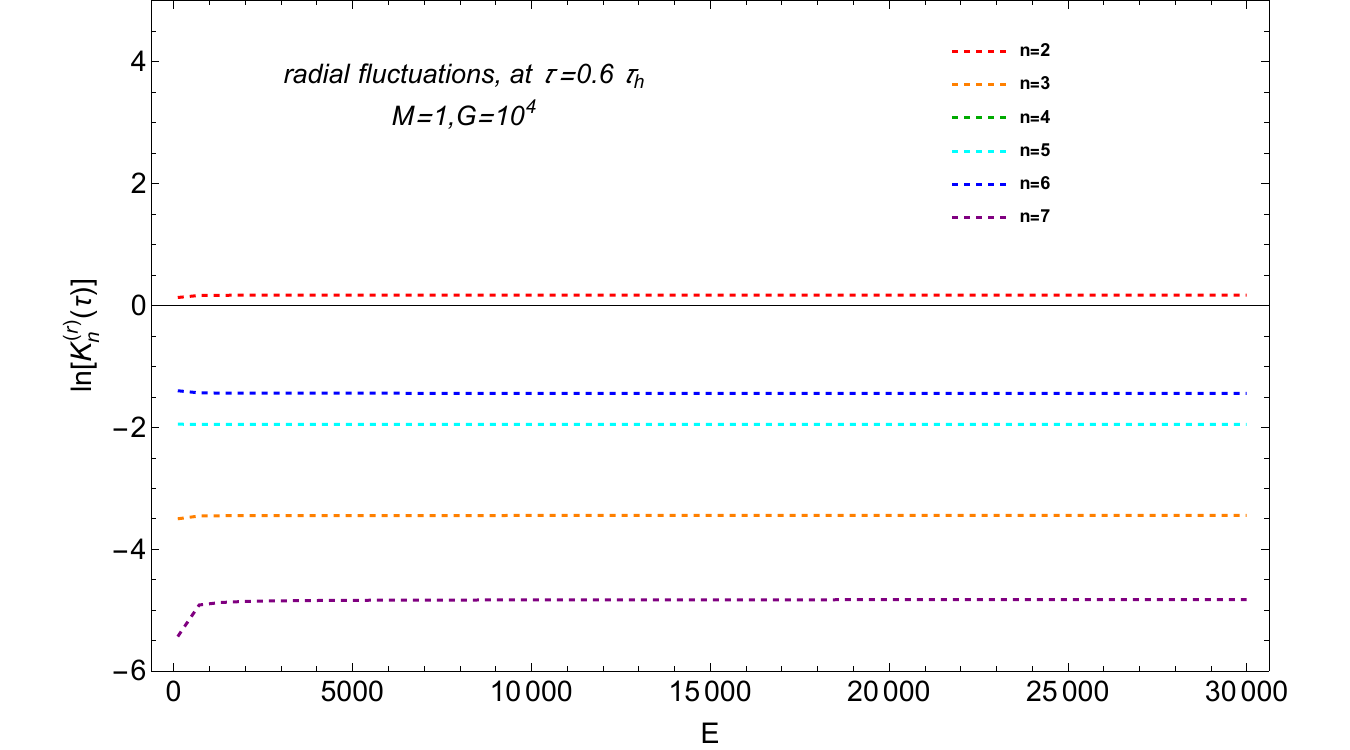}
        \includegraphics[scale=0.5]{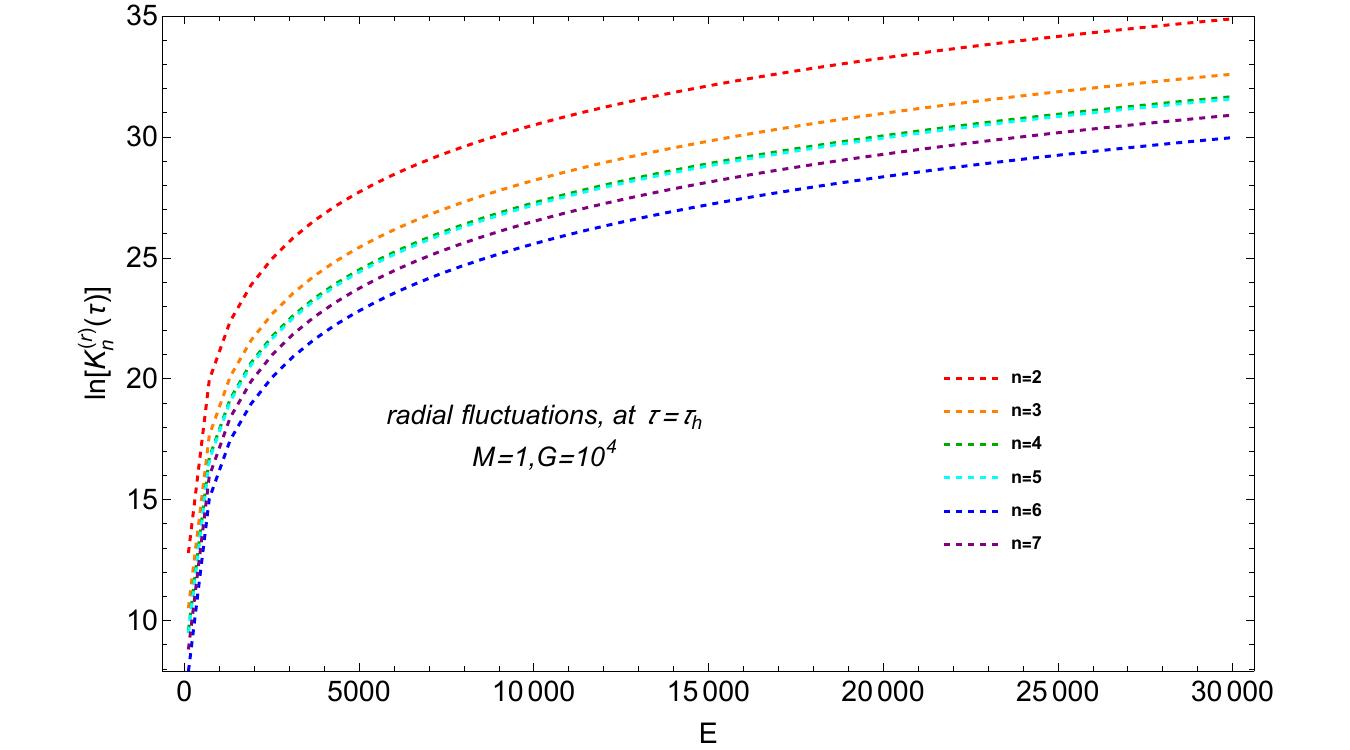}
 		\caption{Krylov complexity, equal to the particle number, associated with radial quantum fluctuations as a function of $E$ at two representative time slices. The upper panel corresponds to an early stage in which the circular string is located far from the black hole horizon, while the lower panel corresponds to a time slice at which the string approaches the vicinity of the event horizon. The vertical axis is displayed on a logarithmic scale, i.e., $\ln(\mathcal{N}_n)$. In the numerical analysis, the parameters are chosen as $\text{G}=10^4$, $M=1$, and $\alpha^\prime=4 M^2_{Pl}=4/\text{G}$, with $E$ explored in the regime $E \gg M$.}
 		\label{ParNumVsETimeSlice}
 	\end{center}
 \end{figure}

 \begin{figure}[H]
 	\begin{center}
 		\includegraphics[scale=0.5]{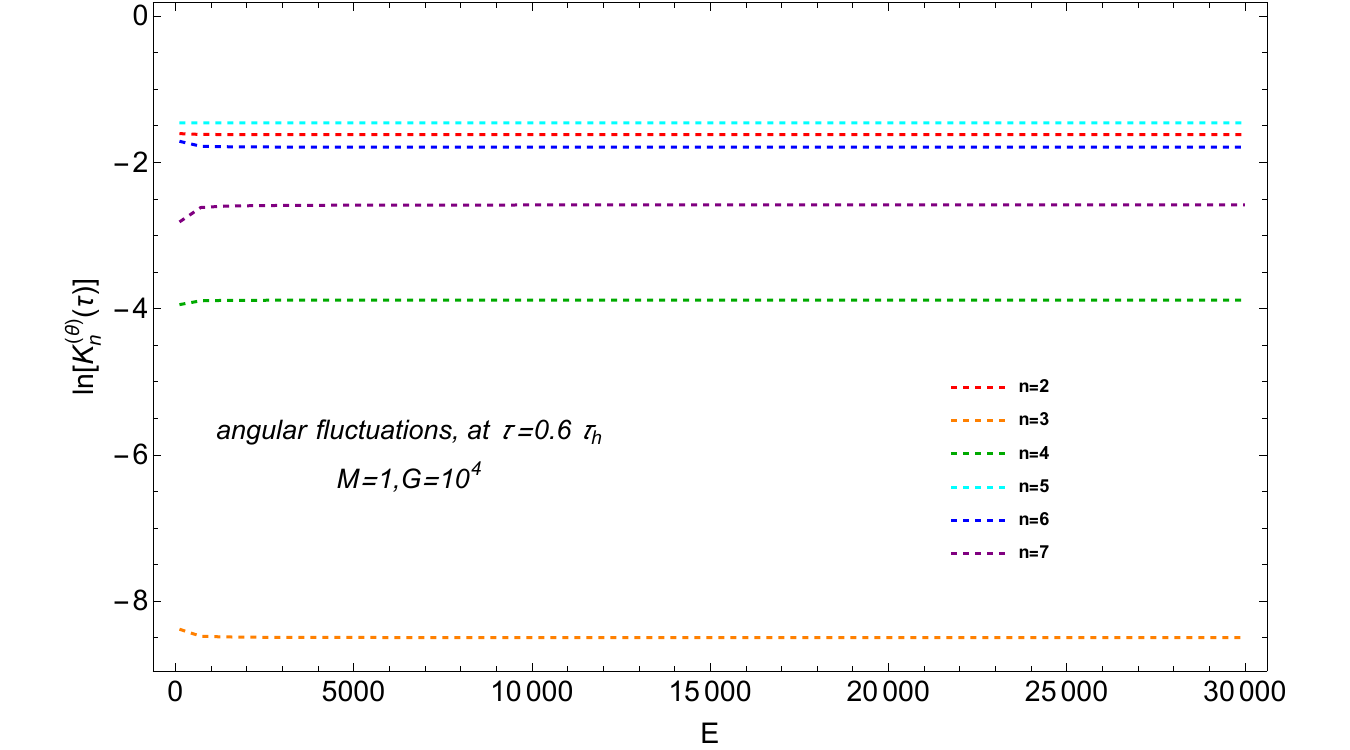}
        \includegraphics[scale=0.5]{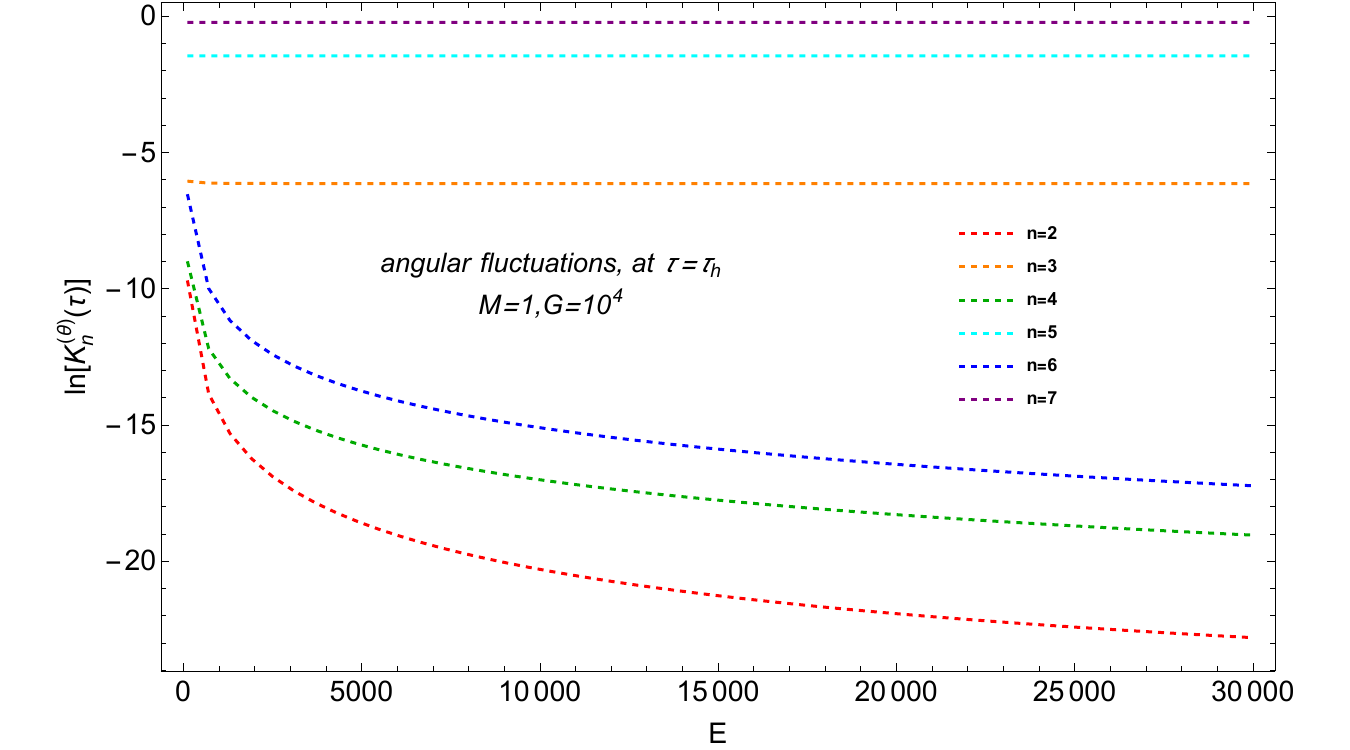}
 		\caption{Krylov complexity associated with angular quantum fluctuations as a function of $E$ at two representative time slices. The parameter choices are identical to those adopted in Fig.~\ref{ParNumVsETimeSlice}.}
 		\label{ParNumVsETimeSliceAngular}
 	\end{center}
 \end{figure}

\section{Krylov Complexity and Lanczos Coefficients Near the Event Horizon \label{ParticleProduction}}

In this section, our goal is to reinterpret the particle number in terms of Krylov complexity and the Lanczos algorithm. For the two-mode quantum state (also referred to as the squeezed-state formalism) \eqref{TwoModeQuantumStateAnyTau}, it has been shown using the Lanczos algorithm \cite{Caputa:2021sib,Bhattacharjee:2022lzy} that the Krylov complexity coincides with the average particle number in each mode \cite{Adhikari:2022oxr}. Accordingly, in the present work, the average particle number in each winding-number mode, namely
\begin{align}
&K^{(\mathtt{i})}_n(\tau)=\langle\Psi_{\gamma,\varphi,\varpi}(\tau)\vert \hat{\mathcal{N}}^{(\mathtt{i})}_n(\tau_0)\vert\Psi_{\gamma,\varphi,\varpi}(\tau)\rangle~,
\end{align}
can equivalently be interpreted as the Krylov complexity of the circular string system. 

Meanwhile, the analysis in \cite{Caputa:2021sib} provides further evidence that Krylov complexity is proportional to the spatial volume, in agreement with the complexity=volume (CV) conjecture proposed in the context of AdS/CFT \cite{Susskind:2014rva}. Motivated by this connection, in this section we investigate how the average particle number evaluated at different time slices depends on the size of the macroscopic string, and interpret the result in terms of Krylov complexity. In particular, since the maximal radius of the string at the initial time is proportional to its energy $E$, we use $E$ as a proxy for the string size and analyze the scaling behavior of $\mathcal{N}^{(\mathtt{i})}_{n}$ with respect to $E$. In addition, we examine another physically relevant quantity, namely the operator growth rate, which is proportional to the Lanczos coefficients, and study its dependence on $E$, especially as the circular string approaches the black hole horizon.

\subsection{Particle number as Krylov complexity}

In the next part, by employing the Lanczos algorithm, in particular the relations \eqref{TimeKrylovOpeSquee}–\eqref{KroComplexEqualToPN}, we will establish the equivalence between the particle number and the Krylov complexity. For the moment, we take this equivalence as given. To characterize the dependence of the particle number on $E$ as the circular string approaches the event horizon, one may adopt two equivalent approaches. One may either solve the mode function $\mathcal{R}_{n}$ from the second-order ODE \eqref{PerLinearModeR} and subsequently compute the particle number using the relations \eqref{SqueeAmpliToModeFunc} and \eqref{DefineParticleNum}, or alternatively, solve the pair of squeezing equations \eqref{StringRPerSqueeAmplitude}–\eqref{StringRPerSqueeAngle} directly and obtain the particle number through $\mathcal{N}_{n}=\sinh^{2}(\gamma_{n})$. Both approaches consistently reproduce the numerical results presented in Fig.\ref{ParNumVsETimeSlice}. More explicitly, we analyze the evolution of $\mathcal{N}_{n}(\tau)$ at two representative time slices. During the infall from asymptotic regions toward the near-horizon zone, $\mathcal{N}_{n}(\tau)$ remains essentially insensitive to the energy $E$. By contrast, as the string approaches the event horizon, i.e. in the limit $\tau \to \tau_h$, the system undergoes effective thermalization, leading to a rapid amplification of particle production. In this regime, the particle number $\mathcal{N}_{n}$ exhibits a polynomial growth with respect to the energy $E$.

Therefore, as illustrated in Fig.~\ref{ParNumVsETimeSlice}, a coherent physical picture emerges: when the circular string approaches sufficiently close to the event horizon, it undergoes a rapid thermalization process and quickly reaches thermal equilibrium at temperature $T = 1/\beta$. During this stage, quantum fluctuations in the radial direction lead to significant particle production in the vicinity of $\tau=\tau_h$, with the emitted bosonic quanta following a Bose–Einstein spectrum characteristic of Hawking radiation. The frequencies of the produced particles, $\omega_n(E)$, display only a weak dependence on the winding number $n$, while they depend primarily on the macroscopic initial configuration of the circular string, parameterized by $\text{G}E$. Correspondingly, the dimensionless combination $\beta \omega_n(E)$ exhibits a polynomial dependence that scales as $\mathcal{O}\left(\frac{M}{E}\right)^\alpha$ with $\alpha>1$.

Moreover, as illustrated in Fig.~\ref{ParNumVsETimeSliceAngular}, the quantum fluctuations in the angular direction do not effectively interact with the thermal bath of the surrounding environment. As a consequence, the corresponding quantum state $\vert \Psi \rangle$ does not lose purity through environmental decoherence \cite{Haque:2021kdm}. In other words, during the infall of the circular string from a distant initial position toward the event horizon, the angular-mode fluctuations remain essentially insensitive to the background thermal bath, thereby preserving their purity; no substantial particle excitation is generated in this sector, and these modes ultimately propagate into the black hole predominantly in a wave‐like manner.

\subsection{Lanczos coefficients $(b_{n}^{(r)})_{\tilde{m}}(\tau)$ and operator growth rate $\tilde{\alpha}_{n}^{(r)}(\tau)$}

\begin{figure}[H]
 	\begin{center}
 		\includegraphics[scale=0.5]{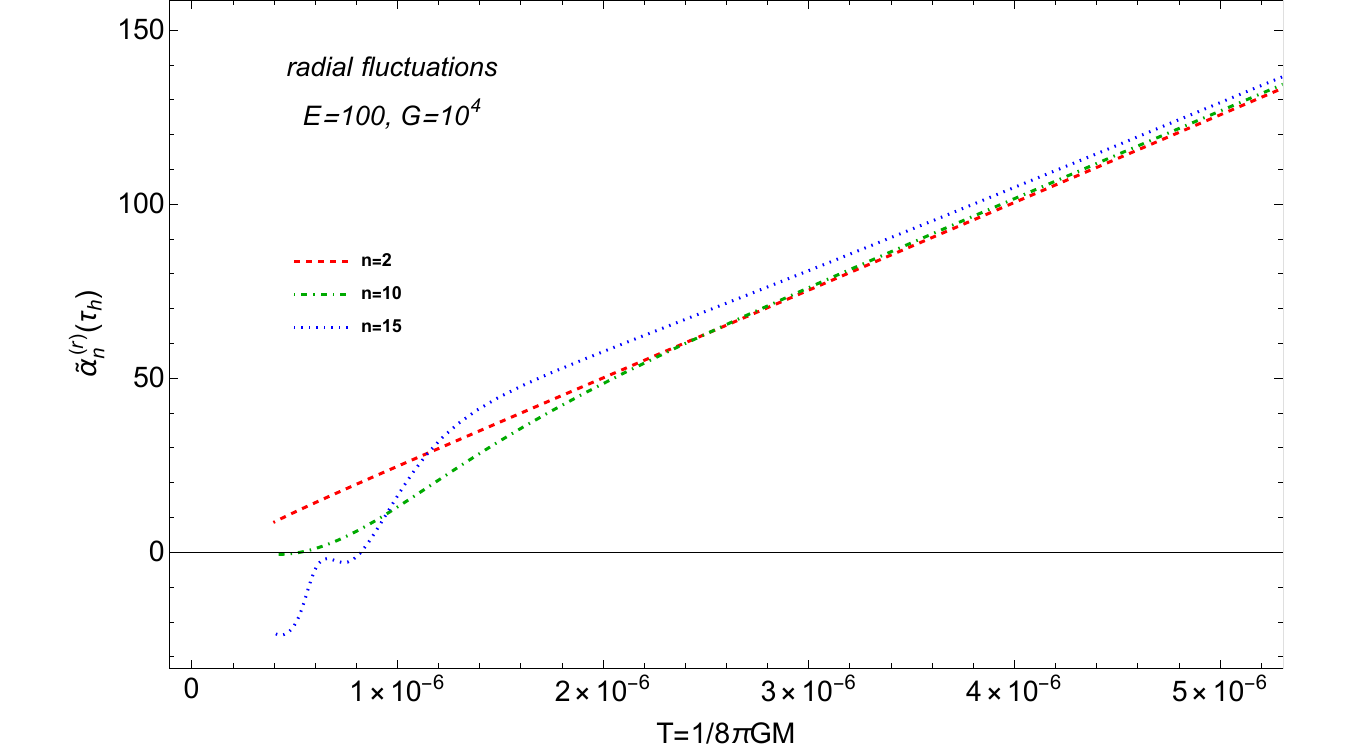}
         \includegraphics[scale=0.5]{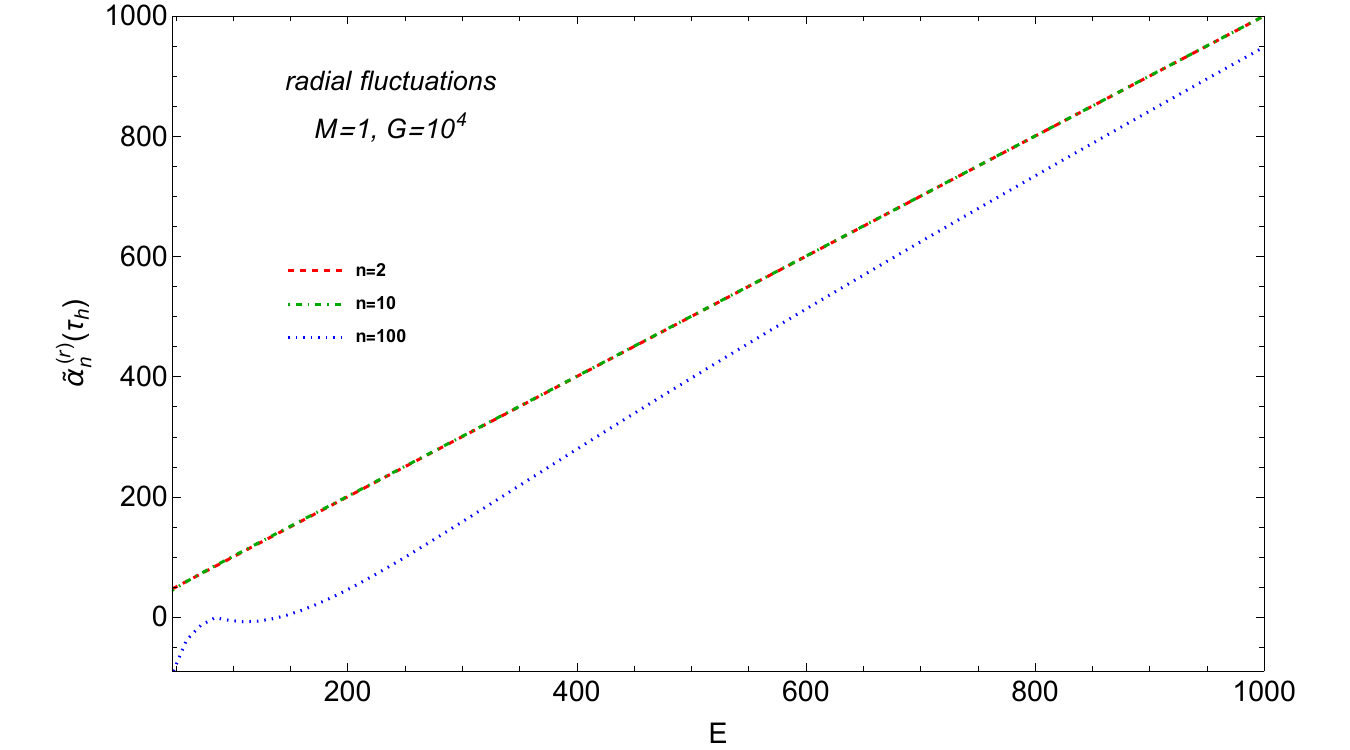}
 	\caption{The operator growth rate $\tilde{\alpha}_{n}^{(r)}(\tau_{h})$, proportional to the Lanczos coefficients $b_{\tilde{m}}(\tau)$, associated with radial quantum fluctuations, shown as a function of the Hawking temperature $T=1/(8\pi \text{G} M)$ (upper panel) and as a function of $E$ (lower panel), both evaluated at the time slice $\tau=\tau_h$, corresponding to the moment when the probe string reaches the black hole horizon.}
 		\label{OPEgrowthVsETimeSliceTauH}
 	\end{center}
 \end{figure}

Subsequently, we briefly outline the construction of Krylov complexity based on the Lanczos algorithm; for further details, see \cite{Parker:2018yvk,Caputa:2021sib}. Regarding the notation used throughout this section, we emphasize that the indices $m,n$ denote the winding numbers appearing in the Fourier expansion of the circular string fluctuations, while $\tilde{m},\tilde{n}$ label the occupation numbers of the Fock states. It should also be noted that quantities without the polarization index $(\mathtt{i})$ and winding number $n$ refer to general definitions, whereas those carrying these indices correspond to the specific setup considered in this work. Given the $SU(1,1)$ symmetry structure of the Hamiltonian \eqref{LieAlgebraFormalismQuadraHamil}, the Liouvillian operator can be constructed as
\begin{align}
\label{GeneLiouvillian}
&\hat{\mathcal{L}}_{n}^{(\mathtt{i})}(\tau)=\tilde{\alpha}_{n}^{(\mathtt{i})}(\tau)(\hat{L}_{n,+}^{(\mathtt{i})}+\hat{L}_{n,-}^{(\mathtt{i})})~,
\end{align}where the coefficient $\tilde{\alpha}_{n}^{(\mathtt{i})}(\tau)$ is a proportionality factor not fixed by symmetry and encodes the operator growth rate. The corresponding raising and lowering operators, $\hat{L}_{n,+}^{(\mathtt{i})}$ and $\hat{L}_{n,-}^{(\mathtt{i})}$, are defined as 
\begin{align}
&\hat{L}_{n,+}^{(\mathtt{i})}(\tau_{0})=-\text{i}\hat{\mathcal{K}}_{n,+}^{(\mathtt{i})}(\tau_{0})=\hat{a}_{n}^{(\mathtt{i})\dagger}(\tau_{0})\hat{a}_{-n}^{(\mathtt{i})\dagger}(\tau_{0})~,\\
&\hat{L}_{n,-}^{(\mathtt{i})}(\tau_{0})=\text{i}\hat{\mathcal{K}}_{n,-}^{(\mathtt{i})}(\tau_{0})=\hat{a}_{-n}^{(\mathtt{i})}(\tau_{0})\hat{a}_{n}^{(\mathtt{i})}(\tau_{0})~.
\end{align}In general, once the Krylov basis $\vert \mathcal{O}_{\tilde{m}}\rangle$ is constructed, the time-evolved operator $\mathcal{O}(\tau)$ can be obtained by acting with the exponential of the Liouvillian on the reference state in the Krylov basis, and can be expanded as
\begin{align}
&\vert\mathcal{O}(\tau)\rangle=\text{e}^{\text{i}\tau\hat{\mathcal{L}}}\vert\mathcal{O}_{0}\rangle=\sum_{\tilde{m}=0}\text{i}^{\tilde{m}}\mathcal{F}_{\tilde{m}}(\tau)\vert\mathcal{O}_{\tilde{m}}\rangle~,
\end{align}where the modulus squared of the coefficients $\mathcal{F}_{\tilde{m}}(\tau)$ defines a probability distribution satisfying the normalization condition
\begin{align}
\label{NormalKrylovOpe}
&\sum_{\tilde{m}}\vert\mathcal{F}_{\tilde{m}}(\tau)\vert^{2}=\sum_{\tilde{m}}P_{\tilde{m}}(\tau)=1~,
\end{align}Based on this probability distribution, the Krylov complexity is defined as
\begin{align}
&K(\tau)=\sum_{\tilde{m}}\tilde{m}P_{\tilde{m}}(\tau)=\sum_{\tilde{m}}\tilde{m}\vert\mathcal{F}_{\tilde{m}}(\tau)\vert^{2}~,
\end{align}In the present setup, the Krylov basis is naturally constructed from the standard two-mode number states in Fock space,
\begin{align}
&\vert(\mathcal{O}_{n}^{(\mathtt{i})})_{\tilde{m}}\rangle=\vert\tilde{m}_{n}^{(\mathtt{i})},\tilde{m}_{-n}^{(\mathtt{i})}\rangle=\frac{1}{\tilde{m}!}\big(\frac{1}{2\pi}\hat{a}_{n}^{(\mathtt{i})\dagger}(\tau_{0})\hat{a}_{-n}^{(\mathtt{i})\dagger}(\tau_{0})\big)^{\tilde{m}}\vert\tilde{0}_{n},\tilde{0}_{-n}\rangle~.
\end{align}By an appropriate choice of the proportionality factor $\tilde{\alpha}_{n}^{(\mathtt{i})}(\tau)$, the action of the exponential of the Liouvillian operator \eqref{GeneLiouvillian} on the ground state of the Krylov basis (i.e. the two-mode vacuum state $\vert\tilde{0}_{n}^{(\mathtt{i})},\tilde{0}_{-n}^{(\mathtt{i})}\rangle$) becomes formally analogous to the action of the squeezing operator $\hat{\mathcal{S}}(\gamma,\varphi)$ on the initial vacuum state, as shown in \eqref{ActingSqueeOnVacuum}. As a consequence, the resulting time-evolved operator can be mapped onto the two-mode quantum state in \eqref{ActingSqueeOnVacuum} with vanishing phase parameter $\varphi^{(\mathtt{i})}_{n}(\tau)$, namely
\begin{align}
\label{TimeKrylovOpeSquee}
&\vert\mathcal{O}_{n}^{(\mathtt{i})}(\tau)\rangle=\sum_{\tilde{m}}\text{i}^{\tilde{m}}\mathcal{F}_{\tilde{m}}(\tau)\vert(\mathcal{O}_{n}^{(\mathtt{i})})_{\tilde{m}}\rangle=\frac{1}{\cosh(\gamma_{n}(\tau))}\sum_{\tilde{m}}\big(\text{i}\,\tanh(\gamma_{n}(\tau))\big)^{\tilde{m}}\vert(\mathcal{O}_{n}^{(\mathtt{i})})_{\tilde{m}}\rangle~.
\end{align}Note that the normalization condition \eqref{NormalKrylovOpe} for the time-evolved operator \eqref{TimeKrylovOpeSquee} is ensured by the series sum rules
\begin{align}
\nonumber
&\frac{1}{\cosh(X)^{2}}\sum_{\tilde{m}=0}^{\infty}\big(\tanh(X)\big)^{2\tilde{m}}=1~.
\end{align}Within the Krylov basis, the Krylov complexity is then defined as
\begin{align}
\label{KroComplexEqualToPN}
&K_{n}^{(\mathtt{i})}(\tau)=\sum_{\tilde{m}}\frac{\tilde{m}}{\cosh(\gamma_{n}^{(\mathtt{i})})^{2}}\big(\tanh(\gamma_{n}^{(\mathtt{i})})\big)^{2\tilde{m}}=\sinh(\gamma_{n}^{(\mathtt{i})})^{2}~.
\end{align}Accordingly, the corresponding Lanczos coefficients $b_{\tilde{m}}(\tau)$ can be identified from the definition
\begin{align}
\label{GeneDefineLanczosCoeff}
&\frac{\partial \mathcal{F}_{\tilde{m}}(\tau)}{\partial \tau}=b_{\tilde{m}}(\tau)\mathcal{F}_{\tilde{m}-1}(\tau)-b_{\tilde{m}+1}(\tau)\mathcal{F}_{\tilde{m}+1}(\tau)~.
\end{align}In the following, we restrict our attention to the radial quantum fluctuations of the string perturbations. This is because the angular fluctuations do not undergo thermalization as the probe string approaches the event horizon, and exhibit no dependence on the energy $E$. Substituting the amplitude $\mathcal{F}_{\tilde{m}}(\tau)$ given in \eqref{TimeKrylovOpeSquee} into the general definition \eqref{GeneDefineLanczosCoeff}, and making use of the equations of motion for $\gamma_{n}^{(r)}(\tau)$ in \eqref{StringRPerSqueeAmplitude}, we obtain
\begin{align}
\label{LancosCoeffString}
&(b_{n}^{(r)})_{\tilde{m}}(\tau)=-\frac{\tilde{m}}{2\pi}\big(n^{2}-\frac{\text{G}M}{\bar{r}(\tau)}-\frac{2\text{G}^{2}E^{2}}{\bar{r}(\tau)^{2}}-\pi^{2}\big)\sin\big(2\varphi_{n}^{(r)}(\tau)\big)~.
\end{align}On the other hand, the Lanczos coefficients $b_{\tilde{m}}(\tau)$ can also be derived by acting with the Liouvillian operator on the Krylov basis,
\begin{align}
\label{LiouviOnKrylov}
&\hat{\mathcal{L}}_{n}^{(r)}(\tau)\vert(\mathcal{O}_{n}^{(\mathtt{i})})_{\tilde{m}}\rangle=(b_{n}^{(r)})_{\tilde{m}}(\tau)\vert(\mathcal{O}_{n}^{(r)})_{\tilde{m}-1}\rangle+(b_{n}^{(r)})_{\tilde{m}+1}(\tau)\vert(\mathcal{O}_{n}^{(r)})_{\tilde{m}+1}\rangle~.
\end{align} By combining \eqref{GeneLiouvillian} with \eqref{LancosCoeffString}–\eqref{LiouviOnKrylov}, the proportionality factor $\tilde{\alpha}_{n}^{(r)}(\tau)$ can be determined as
\begin{align}
\label{OPEgrowthRadialFlu}
&\tilde{\alpha}_{n}^{(r)}(\tau)=\frac{(b_{n}^{(r)})_{\tilde{m}}(\tau)}{2\pi\tilde{m}}=-\frac{1}{4\pi^{2}}\big(n^{2}-\frac{\text{G}M}{\bar{r}(\tau)}-\frac{2\text{G}^{2}E^{2}}{\bar{r}(\tau)^{2}}-\pi^{2}\big)\sin\big(2\varphi_{n}^{(r)}(\tau)\big)~.
\end{align}As indicated in \cite{Parker:2018yvk,Bhattacharjee:2022lzy}, for systems at finite temperature, the operator growth rate $\tilde{\alpha}_{n}^{(r)}$ is bounded by a linear relation proportional to the temperature $T$. This behavior is closely related to the conjectured upper bound on the Lyapunov exponent defined via out-of-time-ordered correlation functions (OTOCs) \cite{Maldacena:2015waa}. In the present setup, as the circular string approaches the black hole horizon and undergoes thermalization, we observe that $\tilde{\alpha}_{n}^{(r)}(\tau_h)$ exhibits a linear scaling with $T$ and remains consistent with this bound. In particular, in the high-temperature regime, $\tilde{\alpha}_{n}^{(r)}(\tau_h)$ approaches and effectively saturates the bound, as illustrated in the upper panel of Fig.~\ref{OPEgrowthVsETimeSliceTauH}. This agreement provides a nontrivial consistency check of the near-horizon dynamics. 

Motivated by the complexity=volume (CV) correspondence for Krylov complexity, we further anticipate a positive correlation between $\tilde{\alpha}_{n}^{(r)}(\tau_h)$ and the string energy $E$. To examine this, we evaluate $\tilde{\alpha}_{n}^{(r)}(\tau_h)$ at the horizon-crossing time $\tau=\tau_h$ and plot it as a function of $E$ for fixed black hole mass $M$, as shown in the lower panel of Fig.~\ref{OPEgrowthVsETimeSliceTauH}. The result exhibits a clear linear growth behavior. This scaling can be understood analytically. In the near-horizon limit $\tau \to \tau_h$, using \eqref{OPEgrowthRadialFlu} together with $\bar{r}(\tau_h)=2\text{G}M$, one finds
\begin{align}
&\tilde{\alpha}_{n}^{(r)}(\tau_{h})=-\frac{1}{4\pi^{2}}\big(n^{2}-\frac{1}{2}-\frac{E^{2}}{2M^{2}}-\pi^{2}\big)\sin\big(2\varphi_{n}^{(r)}(\tau_{h})\big)~,
\end{align}where $\sin\big(2\varphi_{n}^{(r)}(\tau_h)\big) \propto M/E$. This relation directly accounts for the observed linear dependence of $\tilde{\alpha}_{n}^{(r)}(\tau_h)$ on $E$. It should be emphasized, however, that this linear scaling holds only in the regime $n \ll E/(\sqrt{2}M)$. Consistently with \eqref{rbarSol}, we assume $E \gg M$ so that the probe string is initially well separated from the event horizon, ensuring the validity of the near-horizon expansion. In contrast, for large winding numbers $n \gg E/(\sqrt{2}M)$, the behavior of $\tilde{\alpha}_{n}^{(r)}(\tau_h)$ is instead dominated by a contribution scaling as $-M/E$. Nevertheless, modes with large winding number contribute negligibly to particle production compared to the low-$n$ sector. As a result, their impact on the overall dynamics is suppressed, and we restrict our analysis to the regime of small winding numbers.

\section{Conclusion and Discussions \label{ConcluAndDiscuss}}

Inspired by the generalized “complexity = volume” conjecture \cite{Caputa:2021sib,Zhai:2024tkz} in the context of Krylov complexity, and complemented by a wave–particle duality perspective on the wave-spreading analysis of \cite{Larsen:1998sh}, in this work we have investigated particle production, Krylov complexity, and the associated Lanczos coefficients and operator growth rate induced by quantum fluctuations of a circular string contracting toward a black hole horizon, with particular emphasis on the near-horizon regime. In our framework, the particle number is equivalent to the Krylov complexity. Technically, the analysis is formulated within canonical quantization using the squeezed-state formalism. 

Focusing on radial fluctuations, we find that substantial particle production arises only as the circular string approaches the event horizon. This behavior is fully consistent with, and provides a particle-based counterpart to, the suppression of wave-packet spreading in the radial direction reported in \cite{Larsen:1998sh}. Moreover, the dependence of the particle number on the size of the macroscopic string, parametrically controlled by $GE$ and exhibiting a polynomial growth behavior, becomes manifest only once the system enters the near-horizon region and undergoes effective thermalization, as illustrated in Fig.~\ref{ParNumVsETimeSlice}. Equivalently, this indicates that the characteristic scaling associated with the “complexity = volume” conjecture becomes manifest only when the two-mode quantum state describing the fluctuations approaches the thermofield double state.

More intriguingly, in this thermalized near-horizon regime, the operator growth rate $\tilde{\alpha}_{n}^{(r)}$, which is proportional to the Lanczos coefficients and closely related to the conjectured bound on the Lyapunov exponent defined via out-of-time-ordered correlation functions (OTOCs), exhibits a linear dependence on the energy $E$. This behavior provides further support for the proportionality between complexity and the effective size (or volume) of the system. It is worth noting that, for generic finite-temperature systems, the operator growth rate is known to be bounded by a linear relation in the temperature $T$, as illustrated in the upper panel of Fig.~\ref{OPEgrowthVsETimeSliceTauH}. In contrast, in the present circular string–black hole system, we uncover an additional linear scaling with respect to the energy $E$, as shown in the lower panel of Fig.~\ref{OPEgrowthVsETimeSliceTauH}, which can be traced back to the structure implied by the “complexity = volume” correspondence in the Krylov framework. We conjecture that this linear-in-$E$ growth bound near the event horizon may possess a degree of universality. It would be interesting to examine whether a similar behavior persists for other types of horizons, such as trapped or apparent horizons. Furthermore, it remains an open question whether non-singular black hole geometries exhibit an analogous linear scaling of $\tilde{\alpha}_{n}^{(r)}$ with $E$ in the near-horizon region.

In contrast, in the angular direction, as shown in Fig.~\ref{ParNumVsETimeSliceAngular}, particle excitation remains consistently negligible, independent of whether the circular string is far from or near the event horizon. Owing to the absence of significant particle production, the wave packet associated with angular fluctuations preserves its coherence and predominantly wave-like character, ultimately spreading smoothly across the horizon. In other words, the quantum state $\vert \Psi \rangle$ (corresponding to the angular-direction quantum fluctuations) does not lose purity through environmental decoherence. This behavior is fully consistent with the conclusions of \cite{Larsen:1998sh}.

Actually, beyond the notion of Krylov complexity, it has been shown in \cite{Belfiglio:2023moe,Belfiglio:2025ofg,Liu:2026mzz} that there also exist close connections between multipartite entanglement and particle production in an inflationary cosmological background. This observation naturally motivates further investigation in our setup. In future work, it would be interesting to explore whether similar connections between multipartite entanglement and particle production arise in the system of an infalling circular cosmic string. Recent studies of entanglement generated by quantum fluctuations of circular strings have been applied to spacetimes with topological charge \cite{Li:2026rut}. Extending this framework to the present setup may provide further insight into the quantum information–theoretic structure underlying particle creation processes in curved spacetime backgrounds.

In addition, as emphasized in \cite{Das:2025kyn}, the squeezed-state formalism provides a particularly natural and powerful framework for revealing the quantum, or more precisely semi-classical, features of systems with gravitationally induced time dependence, especially when the dynamics is governed by a quadratic-order perturbative action. Such situations arise, for example, when the spacetime curvature itself evolves in time, as in cosmological backgrounds or black-hole merger geometries. Even in flat or stationary spacetimes, time-dependent matter sources whose backreaction is neglected typically give rise to squeezed states once the action is expanded to second order in metric or matter perturbations. From this perspective, the system studied in the present work (the motion of a circular cosmic string in a static Schwarzschild background) falls naturally into the latter category. The spacetime geometry is stationary, while the string provides a dynamical matter source, and the resulting perturbative sector offers a representative physical setting in which the emergence and physical implications of squeezed states can be systematically investigated within a semi-classical framework. And let us look ahead, several promising research directions can be pursued. One possibility is to generalize the present analysis to other stationary but curved backgrounds, such as (A)dS spacetimes, wormhole geometries, or rotating black holes. Another direction is to extend the framework to genuinely time-dependent spacetimes, including FLRW, Vaidya, or Kasner backgrounds, where the interplay between background time dependence, string dynamics, and quantum squeezing is expected to exhibit qualitatively new features.

\section{Acknowledgements}

Ai-chen Li was supported by funding from the China Scholarship Council (CSC) with Grant No.202008620074. This work was also supported by the National Natural Science Foundation of China (NSFC) under Grant No.~12565008, the Youth Program of the Natural Science Foundation of Guangxi under Grant No.~2021GXNSFBA075049.

\appendix

\section{Symmetry structure of the Hamiltonian operator, time-evolution operators, and evolution equations for the squeezing parameters \label{EvolveSqueezingParameter}}

Actually, once the vacuum state at the initial time $\tau_{0}$ is specified, the creation and annihilation operators, i.e., $\hat{a}_{n}^{(r,\theta)\dagger}$ and $\hat{a}_{n}^{(r,\theta)}$ can be expressed in the Heisenberg picture through the Bogoliubov transformation
\begin{align} 
\label{BogoliuAnnihila}
&\hat{a}_{n}^{(\mathtt{i})}(\tau)=U_{n}^{(\mathtt{i})}(\tau)\hat{a}_{n}^{(\mathtt{i})}(\tau_{0})+V_{n}^{(\mathtt{i})}(\tau)\hat{a}_{-n}^{(\mathtt{i})\dagger}(\tau_{0})\\
\label{BogoliuCreation}
&\hat{a}_{-n}^{(\mathtt{i})\dagger}(\tau)=U_{n}^{(\mathtt{i})\star}(\tau)\hat{a}_{-n}^{(\mathtt{i})\dagger}(\tau_{0})+V_{n}^{(\mathtt{i})\star}(\tau)\hat{a}_{n}^{(\mathtt{i})}(\tau_{0})
\end{align}where the winding number satisfies $n \geq 2$. Here, the label $(\mathtt{i})$ denotes the polarization index, corresponding to quantum fluctuations of the probe string in the radial $(r)$ and angular $(\theta)$ directions, respectively. By requiring the preservation of the commutation relations
\begin{align}
&2\pi\delta_{n,l}=[\hat{a}_{n}^{(\mathtt{i})}(\tau),\hat{a}_{l}^{(\mathtt{i})\dagger}(\tau)]=[\hat{a}_{-n}^{(\mathtt{i})}(\tau),\hat{a}_{-l}^{(\mathtt{i})\dagger}(\tau)]
\end{align}the Bogoliubov coefficients $U_{n}^{(\mathtt{i})}(\tau)$ and $V_{n}^{(\mathtt{i})}(\tau)$ should fulfill the following normalization conditions
\begin{align}
&\vert U_{n}^{(\mathtt{i})}(\tau)\vert^{2}-\vert V_{n}^{(\mathtt{i})}(\tau)\vert^{2}=1
\end{align}Moreover, from the perspective of dynamical evolution, the time-dependent operator $\hat{a}_{n}^{(\mathtt{i})}(\tau)$ must satisfy the Heisenberg equation
\begin{align}
\label{HeisenEquaAnnihila}
&\frac{d\hat{a}_{n}^{(\mathtt{i})}(\tau)}{d\tau}=\text{i}[\hat{\mathcal{H}}_{(\mathtt{i})}^{\text{(2)}}(\tau),\hat{a}_{n}^{(\mathtt{i})}(\tau)]
\end{align}After plugging \eqref{HamilQuadraticPer} and \eqref{BogoliuAnnihila} into \eqref{HeisenEquaAnnihila}, the time evolution equations for the Bogoliubov coefficients $U_{n}^{(r)}, V_{n}^{(r)}, U_{n}^{(\theta)}, V_{n}^{(\theta)}$ are found\begin{small}
\begin{align}
\label{EOMsBogoliuUr}
&\frac{d}{d\tau}\!U_{n}^{(r)}(\tau)\!=\!-\frac{\text{i}}{2\pi}\!U_{n}^{(r)}(\tau)(n^{2}\!-\frac{\text{G}\!M}{\bar{r}(\tau)}-\frac{2\text{G}^{2}\!E^{2}}{\bar{r}(\tau)^{2}}+\!\pi^{2})\!-\frac{\text{i}}{2\pi}\!V_{n}^{(r)\star}(\tau)(n^{2}\!-\frac{\text{G}\!M}{\bar{r}(\tau)}-\frac{2\text{G}^{2}\!E^{2}}{\bar{r}(\tau)^{2}}-\!\pi^{2})\\
\label{EOMsBogoliuVr}
&\frac{d}{d\tau}\!V_{n}^{(r)}(\tau)\!=-\frac{\text{i}}{2\pi}\!U_{n}^{(r)\star}(\tau)(n^{2}\!-\frac{\text{G}\!M}{\bar{r}(\tau)}-\frac{2\text{G}^{2}\!E^{2}}{\bar{r}(\tau)^{2}}-\!\pi^{2})\!-\frac{\text{i}}{2\pi}\!V_{n}^{(r)}(\tau)(n^{2}\!-\frac{\text{G}\!M}{\bar{r}(\tau)}-\frac{2\text{G}^{2}\!E^{2}}{\bar{r}(\tau)^{2}}+\!\pi^{2})\\
\label{EOMsBogoliuUtheta}
&\frac{d}{d\tau}U_{n}^{(\theta)}(\tau)=-\frac{\text{i}}{2\pi}U_{n}^{(\theta)}(\tau)(n^{2}-\frac{\text{G}M}{\bar{r}(\tau)}+\pi^{2})-\frac{\text{i}}{2\pi}V_{n}^{(\theta)\star}(\tau)(n^{2}-\frac{\text{G}M}{\bar{r}(\tau)}-\pi^{2})\\
\label{EOMsBogoliuVtheta}
&\frac{d}{d\tau}V_{n}^{(\theta)}(\tau)=-\frac{\text{i}}{2\pi}U_{n}^{(\theta)\star}(\tau)(n^{2}-\frac{\text{G}M}{\bar{r}(\tau)}-\pi^{2})-\frac{\text{i}}{2\pi}V_{n}^{(\theta)}(\tau)(n^{2}-\frac{\text{G}M}{\bar{r}(\tau)}+\pi^{2})
\end{align}
\end{small}In principle, given the quadratic Hamiltonian \eqref{HamilQuadraticPer}, the time evolution operator $\hat{\mathcal{U}}(\tau,\tau_{0})$ can be formally constructed as
\begin{align}
\label{DefineTimeEvoluOperator}
&\hat{\mathcal{U}}(\tau,\tau_{0})=\hat{\mathcal{U}}^{(r)}(\tau,\tau_{0})\hat{\mathcal{U}}^{(\theta)}(\tau,\tau_{0})~,~\hat{\mathcal{U}}^{(\mathtt{i})}(\tau,\tau_{0})=\mathcal{T}\exp\big(-\text{i}\int_{\tau_{0}}^{\tau}d\tilde{\tau}\,\hat{\mathcal{H}}_{(\mathtt{i})}^{\text{(2)}}(\tilde{\tau})\big)
\end{align}However, in practical computations, expanding the exponential of creation and annihilation operators is difficult, as it requires the application of the Zassenhaus formula \cite{Casas:2012nqk,Kimura:2017xxz} along with the evaluation of numerous nested commutators. Alternatively, symmetry considerations and Lie group structures provide a convenient and systematic framework for building the time evolution operator $\hat{\mathcal{U}}(\tau,\tau_{0})$. In terms of the matrix representation, the classical Lie algebra $\mathfrak{su}(1,1)$ is spanned by the generators
\begin{align}
\label{SU11LinearMatrxGenerator}
&\boldsymbol{\mathcal{K}}_{1}=\frac{\text{i}}{2}\sigma_{1}~,~\boldsymbol{\mathcal{K}}_{2}=\frac{\text{i}}{2}\sigma_{2}~,~\boldsymbol{\mathcal{K}}_{3}=\frac{1}{2}\sigma_{3}\\
\label{SU11LinearMatrxGeneratorv1}
&\boldsymbol{\mathcal{K}}_{\pm}=\boldsymbol{\mathcal{K}}_{1}\pm\text{i}\boldsymbol{\mathcal{K}}_{2}~,~\boldsymbol{\mathcal{K}}_{z}=\boldsymbol{\mathcal{K}}_{3}
\end{align}which obey the commutators
\begin{align}
\label{CommutationSU11}&[\boldsymbol{\mathcal{K}}_{+},\boldsymbol{\mathcal{K}}_{-}]=-2\boldsymbol{\mathcal{K}}_{z}~,~[\boldsymbol{\mathcal{K}}_{z},\boldsymbol{\mathcal{K}}_{\pm}]=\pm\boldsymbol{\mathcal{K}}_{\pm}
\end{align}Regarding to the infinitesimal real parameters $\epsilon_{n}^{(a)}$, the transformation matrix associated with the Lie group $SU(1,1)$ takes the form
\begin{align}
&\boldsymbol{\mathcal{M}}_{n}\backsimeq\boldsymbol{I}_{2\times2}+\sum_{a=1}^{3}\text{i}\epsilon_{n}^{(a)}\boldsymbol{\mathcal{K}}_{a}
\end{align}Note that $\hat{\vec{a}}_{n}$ exhibits both vectorial and operator characteristics. In terms of its vectorial nature, it fulfills the canonical transformation relation
\begin{align}
\label{CanonicalTransMatrix}
&\widetilde{\hat{\vec{a}}_{n}^{(\mathtt{i})}}=\boldsymbol{\mathcal{M}}_{n}\cdot\hat{\vec{a}}_{n}^{(\mathtt{i})}\backsimeq\hat{\vec{a}}_{n}^{(\mathtt{i})}+\sum_{a=1}^{3}\text{i}\epsilon_{n}^{(a)}\boldsymbol{\mathcal{K}}_{a}\cdot\hat{\vec{a}}_{n}^{(\mathtt{i})}+\mathcal{O}(\epsilon^{2})
\end{align}On the other hand, as for its operator nature, we promote $\boldsymbol{\mathcal{M}}_{n}\to\hat{\mathcal{M}}^{(\mathtt{i})}$ and suppose
\begin{align}
\label{OPESU11Transform}
&\hat{\mathcal{M}}^{(\mathtt{i})}\backsimeq\hat{\mathbb{I}}+\frac{\text{i}}{2\pi}\sum_{m=2}\sum_{b=1}^{3}\epsilon_{m}^{(b)}\hat{\mathcal{K}}_{b,m}^{(\mathtt{i})}\,,\,\hat{\mathcal{K}}_{b,m}^{(\mathtt{i})}=\hat{\vec{a}}_{m}^{\dagger(\mathtt{i})}\boldsymbol{Q}^{(b)}\hat{\vec{a}}_{m}^{(\mathtt{i})}
\end{align}where the matrix $\boldsymbol{Q}^{(b)}$ remains to be determined. Meanwhile, the unitarity of $\hat{\mathcal{M}}^{(\mathtt{i})}$ imposes the following constraints
\begin{align}
&\hat{\mathcal{K}}_{b,m}^{\dagger,(\mathtt{i})}=\hat{\mathcal{K}}_{b,m}^{(\mathtt{i})}~,~\boldsymbol{Q}^{(b)\dagger}=\boldsymbol{Q}^{(b)}
\end{align}Under the transformation induced by the operator \eqref{OPESU11Transform}, the operator $\hat{\vec{a}}_{n}^{(\mathtt{i})}$ transforms according to the following law
\begin{align}
\label{CanonicalTransOperator}
&\widetilde{\hat{\vec{a}}_{n}^{(\mathtt{i})}}=\hat{\mathcal{M}}^{\dagger(\mathtt{i})}\hat{\vec{a}}_{n}^{(\mathtt{i})}\hat{\mathcal{M}}^{(\mathtt{i})}\backsimeq\hat{\vec{a}}_{n}^{(\mathtt{i})}+\frac{\text{i}}{2\pi}\sum_{m=2}\sum_{b=1}^{3}\epsilon_{m}^{(b)}[\hat{\vec{a}}_{n}^{(\mathtt{i})},\hat{\mathcal{K}}_{b,m}^{(\mathtt{i})}]+\mathcal{O}(\epsilon^{2})
\end{align}By matching \eqref{CanonicalTransMatrix} with \eqref{CanonicalTransOperator}, we could get
\begin{align}
&\text{i}\epsilon_{n}^{(b)}(\boldsymbol{\mathcal{K}}_{b})_{ij_{1}}(\hat{\vec{a}}_{n}^{(\mathtt{i})})_{j_{1}}=\frac{\text{i}}{2\pi}\sum_{m=2}\epsilon_{m}^{(b)}[(\hat{\vec{a}}_{n}^{(\mathtt{i})})_{i},(\hat{\vec{a}}_{m}^{\dagger(\mathtt{i})})_{i_{1}}(\boldsymbol{Q}^{(b)})_{i_{1} j_{1}}(\hat{\vec{a}}_{m}^{(\mathtt{i})})_{j_{1}}]
\end{align}which ultimately leads to
\begin{align}
\label{SolMatrixQ}
&(\boldsymbol{Q}^{(b)})_{j_1 j_{2}}=(\text{i}\boldsymbol{\mathcal{J}_{2}}\boldsymbol{\mathcal{K}}_{b})_{j_1j_{2}}
\end{align}Substituting \eqref{SolMatrixQ} back into \eqref{OPESU11Transform}, the generators $\hat{\mathcal{K}}_{b,n}^{(\mathtt{i})}$ expressed in terms of the creation and annihilation operators are given by
\begin{align}
\label{K1CreationAnnihilation}
&\hat{\mathcal{K}}_{1,n}^{(\mathtt{i})}\!=\!\left(\begin{array}{cc}
\hat{a}_{n}^{(\mathtt{i})\dagger} & \hat{a}_{-n}^{(\mathtt{i})}\end{array}\right)\left(\begin{array}{cc}
0 & \frac{\text{i}}{2}\\
-\frac{\text{i}}{2} & 0
\end{array}\right)\left(\begin{array}{c}
\hat{a}_{n}^{(\mathtt{i})}\\
\hat{a}_{-n}^{(\mathtt{i})\dagger}
\end{array}\right)\!=\!\frac{\text{i}}{2}(\hat{a}_{n}^{(\mathtt{i})\dagger}\hat{a}_{-n}^{(\mathtt{i})\dagger}\!-\!\hat{a}_{-n}^{(\mathtt{i})}\hat{a}_{n}^{(\mathtt{i})})\\
\label{K2CreationAnnihilation}
&\hat{\mathcal{K}}_{2,n}^{(\mathtt{i})}\!=\!\left(\begin{array}{cc}
\hat{a}_{n}^{(\mathtt{i})\dagger} & \hat{a}_{-n}^{(\mathtt{i})}\end{array}\right)\left(\begin{array}{cc}
0 & \frac{1}{2}\\
\frac{1}{2} & 0
\end{array}\right)\left(\begin{array}{c}
\hat{a}_{n}^{(\mathtt{i})}\\
\hat{a}_{-n}^{(\mathtt{i})\dagger}
\end{array}\right)\!=\frac{1}{2}(\hat{a}_{n}^{(\mathtt{i})\dagger}\hat{a}_{-n}^{(\mathtt{i})\dagger}\!+\!\hat{a}_{-n}^{(\mathtt{i})}\hat{a}_{n}^{(\mathtt{i})})\\
\label{K3CreationAnnihilation}
&\hat{\mathcal{K}}_{3,n}^{(\mathtt{i})}\!=\!\left(\begin{array}{cc}
\hat{a}_{n}^{(\mathtt{i})\dagger} & \hat{a}_{-n}^{(\mathtt{i})}\end{array}\right)\left(\begin{array}{cc}
\frac{1}{2} & 0\\
0 & \frac{1}{2}
\end{array}\right)\left(\begin{array}{c}
\hat{a}_{n}^{(\mathtt{i})}\\
\hat{a}_{-n}^{(\mathtt{i})\dagger}
\end{array}\right)\!=\!\frac{1}{2}(\hat{a}_{n}^{(\mathtt{i})\dagger}\hat{a}_{n}^{(\mathtt{i})}+\hat{a}_{-n}^{(\mathtt{i})}\hat{a}_{-n}^{(\mathtt{i})\dagger})
\end{align}Moreover, it is more convenient to define the raising and lowering operators in terms of $\hat{\mathcal{K}}_{1,n}^{(\mathtt{i})}$ and $\hat{\mathcal{K}}_{2,n}^{(\mathtt{i})}$
\begin{align}
\label{KplusCreationAnnihilation}
&\hat{\mathcal{K}}_{+,n}^{(\mathtt{i})}=\hat{\mathcal{K}}_{1,n}^{(\mathtt{i})}+\text{i}\hat{\mathcal{K}}_{2,n}^{(\mathtt{i})}=\text{i}\hat{a}_{n}^{(\mathtt{i})\dagger}(\tau_{0})\hat{a}_{-n}^{(\mathtt{i})\dagger}(\tau_{0})\\
\label{KminusCreationAnnihilation}
&\hat{\mathcal{K}}_{-,n}^{(\mathtt{i})}=\hat{\mathcal{K}}_{1,n}^{(\mathtt{i})}-\text{i}\hat{\mathcal{K}}_{2,n}^{(\mathtt{i})}=-\text{i}\hat{a}_{-n}^{(\mathtt{i})}(\tau_{0})\hat{a}_{n}^{(\mathtt{i})}(\tau_{0})
\end{align}For consistency with standard conventions, we will henceforth denote $\hat{\mathcal{K}}_{3,n}^{(\mathtt{i})}$ by $\hat{\mathcal{K}}_{z,n}^{(\mathtt{i})}$, both in this appendix and throughout the main text. By using \eqref{K1CreationAnnihilation}–\eqref{KminusCreationAnnihilation}, one can recover the commutation relations given in \eqref{CommutationSU11}. More details can be found in Appendix \ref{AppendixNestedCommuta}, where we also introduce several useful nested commutators expressed in terms of the operators \eqref{K1CreationAnnihilation}–\eqref{KminusCreationAnnihilation}. It is now obvious that the Hamiltonian is composed of the generators of the $su(1,1)$ algebra, represented within the framework of quantum group theory. In particular, based on the Bogoliubov transformations \eqref{BogoliuAnnihila}–\eqref{BogoliuCreation}, it can be directly shown that 
\begin{align}
\nonumber
&\hat{a}_{n}^{(\mathtt{i})\dagger}(\tau)\hat{a}_{n}^{(\mathtt{i})}(\tau)+\hat{a}_{-n}^{(\mathtt{i})}(\tau)\hat{a}_{-n}^{(\mathtt{i})\dagger}(\tau)=2\big(\vert U_{n}^{(\mathtt{i})}(\tau)\vert^{2}+\vert V_{n}^{(\mathtt{i})}(\tau)\vert^{2}\big)\hat{\mathcal{K}}_{z,n}^{(\mathtt{i})}(\tau_{0})\\
\label{TimeDependentRotation}
&\quad\quad\quad\quad+2\text{i}U_{n}^{(\mathtt{i})}(\tau)V_{n}^{(\mathtt{i})\star}(\tau)\hat{\mathcal{K}}_{-,n}^{(\mathtt{i})}(\tau_{0})-2\text{i}U_{n}^{(\mathtt{i})\star}(\tau)V_{n}^{(\mathtt{i})}(\tau)\hat{\mathcal{K}}_{+,n}^{(\mathtt{i})}(\tau_{0})
\end{align}
and
\begin{align}
\nonumber
&\hat{a}_{n}^{(\mathtt{i})\dagger}(\tau)\hat{a}_{-n}^{(\mathtt{i})\dagger}(\tau)+\hat{a}_{-n}^{(\mathtt{i})}(\tau)\hat{a}_{n}^{(\mathtt{i})}(\tau)=2\big(U_{n}^{(\mathtt{i})\star}(\tau)V_{n}^{(\mathtt{i})\star}(\tau)+U_{n}^{(\mathtt{i})}(\tau)V_{n}^{(\mathtt{i})}(\tau)\big)\hat{\mathcal{K}}_{z,n}^{(\mathtt{i})}(\tau_{0})\\
\label{TimeDependentSqueezing}
&\quad\quad\quad\quad\quad+\text{i}\big(U_{n}^{(\mathtt{i})}(\tau)^{2}+V_{n}^{(\mathtt{i})\star}(\tau)^{2}\big)\hat{\mathcal{K}}_{-,n}^{(\mathtt{i})}(\tau_{0})-\text{i}\big(U_{n}^{(\mathtt{i})\star}(\tau)^{2}+V_{n}^{(\mathtt{i})}(\tau)^{2}\big)\hat{\mathcal{K}}_{+,n}^{(\mathtt{i})}(\tau_{0})
\end{align}We then proceed by substituting \eqref{TimeDependentRotation}–\eqref{TimeDependentSqueezing} back into \eqref{HamilQuadraticPer}, which leads to the general expression
\begin{align}
\label{LieAlgebraFormalismQuadraHamil}
&\hat{\mathcal{H}}_{\text{(2)}}^{(\mathtt{i})}=\frac{1}{2\pi}\sum_{n=2}\big(\alpha_{z,n}^{(\mathtt{i})}(\tau)\hat{\mathcal{K}}_{z,n}^{(\mathtt{i})}(\tau_{0})+\text{i}\alpha_{-,n}^{(\mathtt{i})}(\tau)\hat{\mathcal{K}}_{-,n}^{(\mathtt{i})}(\tau_{0})-\text{i}\alpha_{+,n}^{(\mathtt{i})}(\tau)\hat{\mathcal{K}}_{+,n}^{(\mathtt{i})}(\tau_{0})\big)
\end{align}Although the time-dependent coefficients $\alpha_{(\pm,z),n}^{(r)}(\tau)$ and $\alpha_{(\pm,z),n}^{(\theta)}(\tau)$ take different explicit forms, both sets of coefficients satisfy
\begin{align}
&\alpha_{z,n}^{(\mathtt{i})\star}(\tau)=\alpha_{z,n}^{(\mathtt{i})}(\tau)\,,\,\alpha_{-,n}^{(\mathtt{i})\star}(\tau)=\alpha_{+,n}^{(\mathtt{i})}(\tau)
\end{align}Therefore, once the Hamiltonian expressed in terms of the Lie algebra formalism, i.e., \eqref{LieAlgebraFormalismQuadraHamil}, is substituted back into the time evolution operator \eqref{DefineTimeEvoluOperator}, it becomes evident that the resulting structure forms a Lie group of type $SU(1,1)$. According to the left-polar decomposition of a $SU(1,1)$ group element, it yields
\begin{align}
\nonumber
\hat{\mathcal{U}}^{(\mathtt{i})}(\tau;\tau_{0})&=\underbrace{\exp\bigg(\frac{1}{2\pi}\sum_{m=2}\big(\xi_{m}^{(\mathtt{i})}(\tau)\hat{\mathcal{K}}_{+,m}^{(\mathtt{i})}(\tau_{0})-\xi_{m}^{(\mathtt{i})\star}(\tau)\hat{\mathcal{K}}_{-,m}^{(\mathtt{i})}(\tau_{0})\big)\bigg)}_{\hat{\mathcal{S}}^{(\mathtt{i})}\big(\gamma(\tau),\varphi(\tau)\big)}\\
\label{TimeEvolutionOpeDecompose}
&\cdot \underbrace{\exp\big(\frac{1}{2\pi}\sum_{m=2}2\text{i}\varpi_{m}^{(\mathtt{i})}(\tau)\hat{\mathcal{K}}_{z,m}^{(\mathtt{i})}(\tau_{0})\big)}_{\hat{\mathcal{R}}^{(\mathtt{i})}\big(\varpi(\tau)\big)}
\end{align}in which the time-dependent parameter $\xi_{m}^{(\mathtt{i})}(\tau)$ is typically expressed as
\begin{align}
&\xi_{m}^{(\mathtt{i})}(\tau)=-\text{i}\gamma_{m}^{(\mathtt{i})}(\tau)\text{e}^{2\text{i}\varphi_{m}^{(\mathtt{i})}(\tau)}
\end{align}In the squeezed-state formalism, $\gamma_{m}^{(\mathtt{i})}(\tau)$ and $\varphi_{m}^{(\mathtt{i})}(\tau)$ denote the squeezing amplitude and squeezing angle, respectively, while $\varpi^{(\mathtt{i})}_{m}(\tau)$ represents the rotation angle. Correspondingly, the operators $\hat{\mathcal{S}}^{(\mathtt{i})}$ and $\hat{\mathcal{R}}^{(\mathtt{i})}$ refer to the two-mode squeezing operator and the rotation operator, respectively. For the squeezing operators, by applying operator ordering theorems, one can decompose them as
\begin{align}
\nonumber
\hat{\mathcal{S}}^{(\mathtt{i})}\big(\gamma(\tau),\varphi(\tau)\big)&=\exp\big\{\frac{1}{2\pi}\sum_{m=2}\big(\text{e}^{2\text{i}\varphi_{m}^{(\mathtt{i})}}\tanh(\gamma_{m}^{(\mathtt{i})})\hat{a}_{m}^{(\mathtt{i})\dagger}(\tau_{0})\hat{a}_{-m}^{(\mathtt{i})\dagger}(\tau_{0})\big)\big\}\\
\nonumber
&\cdot\exp\big\{\frac{1}{2\pi}\sum_{m=2}\big(-\ln(\cosh(\gamma_{m}^{(\mathtt{i})}))\big(\hat{a}_{-m}^{(\mathtt{i})}(\tau_{0})\hat{a}_{-m}^{(\mathtt{i})\dagger}(\tau_{0})+\hat{a}_{m}^{(\mathtt{i})\dagger}(\tau_{0})\hat{a}_{m}^{(\mathtt{i})}(\tau_{0})\big)\big)\big\}\\
\label{SU11OpeSqueezing}
&\cdot\exp\big\{\frac{1}{2\pi}\sum_{m=2}\big(-\text{e}^{-2\text{i}\varphi_{m}^{(\mathtt{i})}}\tanh(\gamma_{m}^{(\mathtt{i})})\hat{a}_{-m}^{(\mathtt{i})}(\tau_{0})\hat{a}_{m}^{(\mathtt{i})}(\tau_{0})\big)\big\}
\end{align}Further details on the derivation of the operator-ordering theorems associated with the Lie group $SU(1,1)$ can be found in Refs.~\cite{Barnett,Grain:2019vnq,Martin:2015qta,Li:2021kfq}. For completeness, we also present a detailed derivation in Appendix~\ref{OpeOrderSU11}. By acting the time evolution operator on the vacuum state at the initial time $\tau_0$, the two-mode quantum state at an arbitrary time $\tau$ is defined as
\begin{align}
\label{GeneDefineSqueezed}
&\vert\Psi_{\gamma,\varphi,\varpi}^{(\mathtt{i})}(\tau)\rangle=\prod_{n=2}^{\infty}\hat{\mathcal{U}}^{(\mathtt{i})}(\tau;\tau_{0})\vert\tilde{0}_{n},\tilde{0}_{-n}\rangle_{\tau_{0}}=\prod_{n=2}^{\infty}\hat{\mathcal{S}}^{(\mathtt{i})}(\gamma,\varphi)\hat{\mathcal{R}}^{(\mathtt{i})}(\varpi)\vert\tilde{0}_{n},\tilde{0}_{-n}\rangle_{\tau_{0}}
\end{align}where the symbols $\tilde{m},\tilde{n}$ denote the occupation numbers of the quantum state, serving to distinguish them from the winding numbers $m,n$ that appear in the Fourier expansions \eqref{PerturRadialFourierExpan}–\eqref{PerturAngularFourierExpan}. When acting the creation and annihilation operators on the occupation number state, the following rule applies
\begin{align}
\label{AnnihiOnNumberState}
&\hat{a}_{\pm l}^{(\mathtt{i})}(\tau_{0})\vert\tilde{m}_{\pm n}^{(\mathtt{i})}\rangle=\delta_{ln}\sqrt{2\pi\tilde{m}}\vert(\tilde{m}-1)_{\pm n}^{(\mathtt{i})}\rangle \\
\label{CreatOnNumberState}
&\hat{a}_{\pm l}^{\dagger(\mathtt{i})}(\tau_{0})\vert\tilde{m}_{\pm n}^{(\mathtt{i})}\rangle=\delta_{ln}\sqrt{2\pi(\tilde{m}+1)}\vert(\tilde{m}+1)_{\pm n}^{(\mathtt{i})}\rangle
\end{align}
From \eqref{AnnihiOnNumberState}-\eqref{CreatOnNumberState}, it is straightforward to derive the relation
\begin{align}
&\frac{\text{i}}{2\pi}\sum_{m=2}\varpi_{m}^{(\mathtt{i})}(\tau)(\hat{a}_{m}^{(\mathtt{i})\dagger}\hat{a}_{m}^{(\mathtt{i})}+\hat{a}_{-m}^{(\mathtt{i})}\hat{a}_{-m}^{(\mathtt{i})\dagger})\vert0_{n},0_{-n}\rangle_{\tau_{0}}=\text{i}\varpi_{n}^{(\mathtt{i})}\vert0_{n},0_{-n}\rangle_{\tau_{0}}
\end{align}This result implies that the rotation operator $\hat{\mathcal{R}}^{(\mathtt{i})}(\varpi)$ contributes merely an overall, physically irrelevant phase factor, namely $\hat{\mathcal{R}}^{(\mathtt{i})}(\varpi)\vert\tilde{0}_{n},\tilde{0}_{-n}\rangle_{\tau_{0}}=\exp\big(\text{i}\varpi^{(\mathtt{i})}_{n}(\tau)\big)\vert0_{n},0_{-n}\rangle_{\tau_{0}}$. Next, we proceed to apply the squeezing operator $\hat{\mathcal{S}}^{(\mathtt{i})}(\gamma,\varphi)$ to the initial vacuum state. After lengthy but straightforward calculations, we eventually arrive at
\begin{align}
\nonumber
\hat{\mathcal{S}}^{(\mathtt{i})}\vert\tilde{0}_{n},\tilde{0}_{-n}\rangle_{\tau_{0}}&=\!\frac{1}{\cosh(\gamma_{n}^{(\mathtt{i})}(\tau))}\exp\big\{\frac{1}{2\pi}\sum_{m=2}\big(\text{e}^{2\text{i}\varphi_{m}^{(\mathtt{i})}(\tau)}\tanh(\gamma_{m}^{(\mathtt{i})}(\tau))\hat{a}_{m}^{(\mathtt{i})\dagger}(\tau_{0})\hat{a}_{-m}^{(\mathtt{i})\dagger}(\tau_{0})\big)\big\}\vert\tilde{0}_{n},\tilde{0}_{-n}\rangle_{\tau_{0}}\\
\label{ActingSqueeOnVacuum}
&=\frac{1}{\cosh(\gamma_{n}^{(\mathtt{i})}(\tau))}\sum_{\tilde{l}=0}^{\infty}\big(\text{e}^{2\text{i}\varphi_{n}^{(\mathtt{i})}(\tau)}\tanh(\gamma_{n}^{(\mathtt{i})}(\tau))\big)^{\tilde{l}}\vert\tilde{l}_{n}^{(\mathtt{i})},\tilde{l}_{-n}^{(\mathtt{i})}\rangle_{\tau_{0}}
\end{align}in which it is necessary to employ the following iterative relations based on \eqref{AnnihiOnNumberState}-\eqref{CreatOnNumberState}
\begin{align}
\nonumber
&\frac{1}{2\pi}\sum_{m=2}\text{e}^{2\text{i}\varphi_{m}^{(\mathtt{i})}}\tanh(\gamma_{m}^{(\mathtt{i})})\,\hat{a}_{m}^{(\mathtt{i})\dagger}(\tau_{0})\hat{a}_{-m}^{(\mathtt{i})\dagger}(\tau_{0})\vert\tilde{0}_{n},\tilde{0}_{-n}\rangle_{\tau_{0}}=\text{e}^{2\text{i}\varphi_{n}^{(\mathtt{i})}}\tanh\gamma_{n}^{(\mathtt{i})}\,\vert\tilde{1}_{n}^{(\mathtt{i})},\tilde{1}_{-n}^{(\mathtt{i})}\rangle_{\tau_{0}}\\
\nonumber
& \quad\quad\quad\quad\quad\quad\quad\quad\quad\quad\quad\quad\quad\quad \dots\longrightarrow \dots \\
\nonumber
&\big(\frac{1}{2\pi}\sum_{m=2}\text{e}^{2\text{i}\varphi_{m}^{(\mathtt{i})}}\tanh(\gamma_{m}^{(\mathtt{i})})\,\hat{a}_{m}^{(\mathtt{i})\dagger}(\tau_{0})\hat{a}_{-m}^{(\mathtt{i})\dagger}(\tau_{0})\big)^{\tilde{l}}\vert\tilde{0}_{n},\tilde{0}_{-n}\rangle_{\tau_{0}}=\tilde{l}!\,(\text{e}^{2\text{i}\varphi_{n}^{(\mathtt{i})}}\tanh(\gamma_{n}^{(\mathtt{i})}))^{\tilde{l}}\vert\tilde{l}_{n}^{(\mathtt{i})},\tilde{l}_{-n}^{(\mathtt{i})}\rangle_{\tau_{0}}
\end{align}Eventually, the two-mode quantum state at an arbitrary time $\tau$ with winding number $m$ takes the form
\begin{align}
\label{TwoModeQuantumStateAnyTau}
&\vert\Psi_{\gamma,\varphi,\varpi}^{(\mathtt{i})}(\tau)\rangle_{m}=\frac{\text{e}^{\text{i}\varpi_{m}^{(\mathtt{i})}(\tau)}}{\cosh(\gamma_{m}^{(\mathtt{i})}(\tau))}\sum_{\tilde{n}=0}^{\infty}\big(\text{e}^{2\text{i}\varphi_{m}^{(\mathtt{i})}(\tau)}\tanh(\gamma_{m}^{(\mathtt{i})}(\tau))\big)^{\tilde{n}}\vert\tilde{n}_{m}^{(\mathtt{i})},\tilde{n}_{-m}^{(\mathtt{i})}\rangle_{\tau_{0}}
\end{align}On the other hand, with the aid of the time-evolution operator $\hat{\mathcal{U}}^{(\mathtt{i})}(\tau;\tau_{0})$, the creation and annihilation operators at an arbitrary time $\tau$, namely $\hat{a}^{(\mathtt{i})}_{n}(\tau)$ and $\hat{a}_{n}^{(\mathtt{i})\dagger}(\tau)$, can be constructed via
\begin{align}
\label{ConstruAnnihilaOpeAnyTau}
&\hat{a}^{(\mathtt{i})}_{\pm n}(\tau)=\hat{\mathcal{R}}^{(\mathtt{i})\dagger}(\varpi)\hat{\mathcal{S}}^{(\mathtt{i})\dagger}(\gamma,\varphi)\hat{a}^{(\mathtt{i})}_{\pm n}(\tau_{0})\hat{\mathcal{S}}^{(\mathtt{i})}(\gamma,\varphi)\hat{\mathcal{R}}^{(\mathtt{i})}(\varpi)~,~n\geq 2\\
\label{ConstruCreationOpeAnyTau}
&\hat{a}_{\pm n}^{(\mathtt{i})\dagger}(\tau)=\hat{\mathcal{R}}^{(\mathtt{i})\dagger}(\varpi)\hat{\mathcal{S}}^{(\mathtt{i})\dagger}(\gamma,\varphi)\hat{a}_{\pm n}^{(\mathtt{i})\dagger}(\tau_{0})\hat{\mathcal{S}}^{(\mathtt{i})}(\gamma,\varphi)\hat{\mathcal{R}}^{(\mathtt{i})}(\varpi)~,~n\geq 2
\end{align}
In appendix \ref{AppendixNestedCommuta}, we present the detailed derivations for evaluating \eqref{ConstruAnnihilaOpeAnyTau}–\eqref{ConstruCreationOpeAnyTau}. Here, we simply summarize the main results.
\begin{align}
&\hat{a}_{n}^{(\mathtt{i})}(\tau)=\cosh(\gamma_{n}^{(\mathtt{i})}(\tau))\text{e}^{\text{i}\varpi_{n}^{(\mathtt{i})}(\tau)}\hat{a}_{n}^{(\mathtt{i})}(\tau_{0})+\text{e}^{-\text{i}\varpi_{n}^{(\mathtt{i})}(\tau)}\text{e}^{2\text{i}\varphi_{n}^{(\mathtt{i})}(\tau)}\sinh(\gamma_{n}^{(\mathtt{i})}(\tau))\hat{a}_{-n}^{(\mathtt{i})\dagger}(\tau_{0})\\
&\hat{a}_{n}^{\dagger(\mathtt{i})}(\tau)=\cosh(\gamma_{n}^{(\mathtt{i})}(\tau))\text{e}^{-\text{i}\varpi_{n}^{(\mathtt{i})}(\tau)}\hat{a}_{n}^{\dagger(\mathtt{i})}(\tau_{0})+\text{e}^{\text{i}\varpi_{n}^{(\mathtt{i})}(\tau)}\text{e}^{-2\text{i}\varphi_{n}^{(\mathtt{i})}(\tau)}\sinh(\gamma_{n}^{(\mathtt{i})}(\tau))\hat{a}_{-n}^{(\mathtt{i})}(\tau_{0})
\end{align}By matching them with the Bogoliubov transformation \eqref{BogoliuAnnihila}, one can get
\begin{align}
\label{BogoliuToSqueezingPara}
&U_{n}^{(\mathtt{i})}(\tau)=\cosh(\gamma_{n}^{(\mathtt{i})}(\tau))\text{e}^{\text{i}\varpi_{n}^{(\mathtt{i})}(\tau)}~,~V_{n}^{(\mathtt{i})}(\tau)=\text{e}^{2\text{i}\varphi_{n}^{(\mathtt{i})}(\tau)-\text{i}\varpi_{n}^{(\mathtt{i})}(\tau)}\sinh(\gamma_{n}^{(\mathtt{i})}(\tau))~,~n\geq2
\end{align}After substituting them into the evolution equations \eqref{EOMsBogoliuUr}–\eqref{EOMsBogoliuVtheta}, the time evolution of the squeezing parameters corresponding to the radial and angular quantum fluctuations of the macroscopic string can be described by the following two sets of differential equations, respectively.
\begin{align}
\label{StringRPerSqueeAmplitude}
\dot{\gamma}_{n}^{(r)}=&-\frac{1}{2\pi}\sin(2\varphi_{n}^{(r)})(n^{2}-\frac{\text{G}M}{r(\tau)}-\frac{2\text{G}^{2}E^{2}}{r(\tau)^{2}}-\pi^{2})\\
\nonumber
\dot{\varphi}_{n}^{(r)}=&-\frac{\cos(2\varphi_{n}^{(r)})\coth(2\gamma_{n}^{(r)})}{2\pi}(n^{2}-\frac{\text{G}M}{r(\tau)}-\frac{2\text{G}^{2}E^{2}}{r(\tau)^{2}}-\pi^{2})\\
\label{StringRPerSqueeAngle}
&-\frac{1}{2\pi}(n^{2}-\frac{\text{G}M}{r(\tau)}-\frac{2\text{G}^{2}E^{2}}{r(\tau)^{2}}+\pi^{2})\\
\nonumber
\dot{\varpi}_{n}^{(r)}=&-\frac{\cos(2\varphi_{n}^{(r)})\tanh(\gamma_{n}^{(r)})}{2\pi}(n^{2}-\frac{\text{G}M}{r(\tau)}-\frac{2\text{G}^{2}E^{2}}{r(\tau)^{2}}-\pi^{2})\\
\label{StringRPerRotationAngle}
&-\frac{1}{2\pi}(n^{2}-\frac{\text{G}M}{r(\tau)}-\frac{2\text{G}^{2}E^{2}}{r(\tau)^{2}}+\pi^{2})
\end{align}and
\begin{align}
&\dot{\gamma}_{n}^{(\theta)}=-\frac{1}{2\pi}\sin(2\varphi_{n}^{(\theta)})(n^{2}-\frac{\text{G}M}{r(\tau)}-\pi^{2})\\
&\dot{\varphi}_{n}^{(\theta)}=-\frac{1}{2\pi}(n^{2}-\frac{\text{G}M}{r(\tau)}+\pi^{2})-\frac{\cos(2\varphi_{n}^{(\theta)})\coth(2\gamma_{n}^{(\theta)})}{2\pi}(n^{2}-\frac{\text{G}M}{r(\tau)}-\pi^{2})\\
&\dot{\varpi}_{n}^{(\theta)}=-\frac{1}{2\pi}(n^{2}-\frac{\text{G}M}{r(\tau)}+\pi^{2})-\frac{\cos(2\varphi_{n}^{(\theta)})\tanh(\gamma_{n}^{(\theta)})}{2\pi}(n^{2}-\frac{\text{G}M}{r(\tau)}-\pi^{2})
\end{align}Note that the differential equations governing the variables $\gamma^{(\mathtt{i})}_{n}$ and $\varphi^{(\mathtt{i})}_{n}$ are mutually coupled and form a closed system. Once $\gamma^{(\mathtt{i})}_{n}$ and $\varphi^{(\mathtt{i})}_{n}$ are determined, the time evolution of the rotation angle $\varpi^{(\mathtt{i})}_{n}$ is also fixed. This is expected, as $\varpi^{(\mathtt{i})}_{n}$ appears only as a global phase in the two-mode quantum state \eqref{TwoModeQuantumStateAnyTau} and does not contribute to any physical observable of the form $\langle\Psi_{\gamma,\varphi,\varpi}(\tau)\vert\hat{O}\vert\Psi_{\gamma,\varphi,\varpi}(\tau)\rangle$. Consequently, throughout most of the discussion and analysis in this work, we will omit $\varpi^{(\mathtt{i})}_{n}$ without loss of generality.

\section{Useful nested commutators and the action of time-evolution operators on creation and annihilation operators\label{AppendixNestedCommuta}}

In Appendix~\ref{EvolveSqueezingParameter}, we constructed the $su(1,1)$ generators in terms of creation and annihilation operators via the Schwinger boson representation, as given in \eqref{K1CreationAnnihilation}–\eqref{KminusCreationAnnihilation}. In this part, we begin by verifying that these generators satisfy the standard commutation relations of the $su(1,1)$ Lie algebra as a preliminary consistency check, namely
\begin{align}
&\hat{\mathcal{K}}_{\pm}^{(\mathtt{i})}=\frac{1}{2\pi}\sum_{n=2}\hat{\mathcal{K}}_{n,\pm}^{(\mathtt{i})}~,~\hat{\mathcal{K}}_{3}^{(\mathtt{i})}=\frac{1}{2\pi}\sum_{n=2}\hat{\mathcal{K}}_{n,3}^{(\mathtt{i})}
\end{align}and
\begin{align}
\nonumber
[\hat{\mathcal{K}}_{+}^{(\mathtt{i})},\hat{\mathcal{K}}_{-}^{(\mathtt{i})}]&=[\frac{\text{i}}{2\pi}\sum_{n=2}\hat{a}_{n}^{(\mathtt{i})\dagger}(\tau_{0})\hat{a}_{-n}^{(\mathtt{i})\dagger}(\tau_{0}),\frac{-\text{i}}{2\pi}\sum_{l=2}\hat{a}_{-l}^{(\mathtt{i})}(\tau_{0})\hat{a}_{l}^{(\mathtt{i})}(\tau_{0})]\\
\nonumber
&=\frac{1}{(2\pi)^{2}}\sum_{n=2}\sum_{l=2}\big(-\hat{a}_{n}^{(\mathtt{i})\dagger}(\tau_{0})\delta_{n,l}\hat{a}_{l}^{(\mathtt{i})}(\tau_{0})-\hat{a}_{-l}^{(\mathtt{i})}(\tau_{0})\delta_{nl}\hat{a}_{-n}^{(\mathtt{i})\dagger}(\tau_{0})\big)\\
\nonumber
&=-2\times\frac{1}{2\pi}\sum_{n=2}\hat{\mathcal{K}}_{n,3}^{(\mathtt{i})}=-2\hat{\mathcal{K}}_{3}^{(\mathtt{i})}
\end{align}Following the same procedure, one can similarly verify
\begin{align}
\nonumber
&[\hat{\mathcal{K}}_{3}^{(\mathtt{i})},\hat{\mathcal{K}}_{+}^{(\mathtt{i})}]\!=\!+\hat{\mathcal{K}}_{\pm}^{(\mathtt{i})}\,,\,[\hat{\mathcal{K}}_{3}^{(\mathtt{i})},\hat{\mathcal{K}}_{-}^{(\mathtt{i})}]\!=\!-\hat{\mathcal{K}}_{-}^{(\mathtt{i})}
\end{align}
To facilitate the derivation of $\hat{\mathcal{U}}^{\dagger}(\tau;\tau_{0})\hat{a}_{n}(\tau_{0})\hat{\mathcal{U}}(\tau;\tau_{0})\,,\,n\geq 2$, which involves a variety of commutators and operator products, we first present a compilation of frequently used commutation relations among operators in the irreducible Schwinger boson representation, namely
\begin{small}
\begin{align}
\label{CommutatorOneFold}
&[-\frac{1}{2\pi}\sum_{m=2}X_{m}(\tau)\hat{a}_{m}^{\dagger}(\tau_{0})\hat{a}_{-m}^{\dagger}(\tau_{0}),\hat{a}_{n}(\tau_{0})]=X_{n}(\tau)\hat{a}_{-n}^{\dagger}(\tau_{0})\\
&[\sum_{m=2}\frac{X_{m}}{2\pi}\,\hat{a}_{-m}(\tau_{0})\hat{a}_{m}(\tau_{0}),\hat{a}_{-n}^{\dagger}(\tau_{0})]=X_{n}\hat{a}_{n}(\tau_{0})\\
&[\sum_{m_{k}=2}\frac{X_{m_{k}}}{2\pi}\hat{a}_{m_{k}}^{\dagger}(\tau_{0})\hat{a}_{-m_{k}}^{\dagger}(\tau_{0}),\dots[\sum_{m_{1}=2}\frac{X_{m_{1}}}{2\pi}\hat{a}_{m_{1}}^{\dagger}(\tau_{0})\hat{a}_{-m_{1}}^{\dagger}(\tau_{0}),\hat{a}_{n}(\tau_{0})]\dots]\!=\!0, k\geq 2 \\
&[\sum_{m_{k}=2}\frac{X_{m_{k}}}{2\pi}\hat{a}_{-m_{k}}(\tau_{0})\hat{a}_{m_{k}}(\tau_{0}),\dots[\sum_{m_{1}=2}\frac{X_{m_{1}}}{2\pi}\hat{a}_{-m_{1}}(\tau_{0})\hat{a}_{m_{1}}(\tau_{0}),\hat{a}_{-n}^{\dagger}(\tau_{0})]\dots]\!=\!0\, ,\, k\geq2 \\
&[\sum_{m_{k}=2}\!\frac{X_{m_{k}}}{2\pi}\big(\!\hat{a}_{-m_{k}}^{(0)}\hat{a}_{-m_{k}}^{\dagger(0)}\!+\!\hat{a}_{m_{k}}^{\dagger(0)}\hat{a}_{m_{k}}^{(0)}\big),\dots[\sum_{m_{1}=2}\!\frac{X_{m_{1}}}{2\pi}\big(\hat{a}_{-m_{1}}^{(0)}\hat{a}_{-m_{1}}^{\dagger(0)}\!+\!\hat{a}_{m_{1}}^{\dagger(0)}\hat{a}_{m_{1}}^{(0)}\big),\hat{a}_{n}^{(0)}]\dots]\!=\!(-X_{n})^{k}\hat{a}_{n}^{(0)}\\
&[\!\sum_{m_{k}=2}\!\!\!\frac{X_{m_{k}}}{2\pi}\big(\hat{a}_{-m_{k}}^{(0)}\hat{a}_{-m_{k}}^{\dagger(0)}\!+\!\hat{a}_{m_{k}}^{\dagger(0)}\hat{a}_{m_{k}}^{(0)}\big),\dots[\!\sum_{m_{1}=2}\!\!\!\frac{X_{m_{1}}}{2\pi}\big(\hat{a}_{-m_{1}}^{(0)}\hat{a}_{-m_{1}}^{\dagger(0)}\!+\!\hat{a}_{m_{1}}^{\dagger(0)}\hat{a}_{m_{1}}^{(0)}\big),\hat{a}_{-n}^{\dagger}(\tau_{0})]\dots]\!=\!(X_{n})^{k}\hat{a}_{-n}^{\dagger(0)}\\
\label{CommutatorLastExpV1}
&\exp\big\{\!\!\sum_{m=2}\!\!\frac{X_{m}}{2\pi}(\hat{a}_{-m}^{(0)}\hat{a}_{-m}^{\dagger(0)}\!+\!\hat{a}_{m}^{\dagger(0)}\hat{a}_{m}^{(0)})\big\}\cdot\hat{a}_{n}^{(0)}\cdot\exp\big\{\!-\!\sum_{m=2}\!\!\frac{X_{m}}{2\pi}(\hat{a}_{-m}^{(0)}\hat{a}_{-m}^{\dagger(0)}\!+\!\hat{a}_{m}^{\dagger(0)}\hat{a}_{m}^{(0)})\big\}\!=\!\text{e}^{-X_{n}}\hat{a}_{n}^{(0)}\\
\label{CommutatorLastExpV2}
&\exp\big\{\!\!\sum_{m=1}\!\!\frac{X_{m}}{2\pi}(\hat{a}_{-m}^{(0)}\hat{a}_{-m}^{\dagger(0)}\!+\!\hat{a}_{m}^{\dagger(0)}\hat{a}_{m}^{(0)})\big\}\cdot\hat{a}_{-n}^{\dagger(0)}\cdot\exp\big\{\!-\!\sum_{m=1}\!\!\frac{X_{m}}{2\pi}(\hat{a}_{-m}^{(0)}\hat{a}_{-m}^{\dagger(0)}\!+\!\hat{a}_{m}^{\dagger(0)}\hat{a}_{m}^{(0)})\big\}\!=\!\text{e}^{X_{n}}\hat{a}_{-n}^{\dagger(0)}
\end{align}
\end{small}By employing the commutator identities \eqref{CommutatorOneFold}–\eqref{CommutatorLastExpV2} for $\hat{\mathcal{S}}^{\dagger}\hat{a}_{n}(\tau_{0})\hat{\mathcal{S}}$, and the identities \eqref{CommutatorLastExpV1}-\eqref{CommutatorLastExpV2} for $\hat{\mathcal{R}}^{\dagger}\hat{\mathcal{S}}^{\dagger}\hat{a}_{n}(\tau_{0})\hat{\mathcal{S}}\hat{\mathcal{R}}$, we arrive at
\begin{align}
\label{SqueezingOpeOnAnnihilation}
&\hat{\mathcal{S}}^{\dagger}(\gamma,\varphi)\hat{a}_{n}(\tau_{0})\hat{\mathcal{S}}(\gamma,\varphi)=\cosh(\gamma_{n})\hat{a}_{n}(\tau_{0})+\text{e}^{2\text{i}\varphi_{n}}\sinh(\gamma_{n})\hat{a}_{-n}^{\dagger}(\tau_{0})\\
\label{RotationOpeOnAnnihilation}
&\hat{\mathcal{R}}^{\dagger}(\varpi)\hat{a}_{n}(\tau_{0})\hat{\mathcal{R}}(\varpi)=\text{e}^{\text{i}\varpi_{n}}\hat{a}_{n}^{(0)}\,,\,\hat{\mathcal{R}}^{\dagger}(\varpi)\hat{a}_{-n}^{\dagger}(\tau_{0})\hat{\mathcal{R}}(\varpi)=\text{e}^{-\text{i}\varpi_{n}}\hat{a}_{-n}^{\dagger(0)}
\end{align}Eventually, on the basis of the results \eqref{SqueezingOpeOnAnnihilation}-\eqref{RotationOpeOnAnnihilation} and the decomposition law \eqref{TimeEvolutionOpeDecompose}, the time evolution of the annihilation operator is governed by
\begin{align}
\nonumber
\hat{a}_{n}(\tau)&=\!\hat{\mathcal{U}}^{\dagger}(\tau;\tau_{0})\hat{a}_{n}(\tau_{0})\hat{\mathcal{U}}(\tau;\tau_{0})\\
\label{TimeEvolutionAnnihi}
&=\cosh(\gamma_{n}(\tau))\text{e}^{\text{i}\varpi_{n}(\tau)}\hat{a}_{n}(\tau_{0})+\text{e}^{2\text{i}\varphi_{n}(\tau)-\text{i}\varpi_{n}(\tau)}\sinh(\gamma_{n}(\tau))\hat{a}_{-n}^{\dagger}(\tau_{0})
\end{align}Following the same procedure, we could also obtain
\begin{align}
\label{TimeEvolutionCreation}
&\hat{a}_{n}^{\dagger}(\tau)=\text{e}^{-2\text{i}\varphi_{n}(\tau)+\text{i}\varpi_{n}(\tau)}\sinh(\gamma_{n}(\tau))\hat{a}_{-n}(\tau_{0})+\cosh(\gamma_{n}(\tau))\text{e}^{-\text{i}\varpi_{n}(\tau)}\hat{a}_{n}^{\dagger}(\tau_{0})
\end{align}

\section{Operator-ordering theorems for the Lie group $SU(1,1)$ \label{OpeOrderSU11}}

We begin by introducing the generators of the Lie algebra $su(1,1)$, namely \begin{align}
\label{liealgebrasu11}
&[\hat{K}_{3},\hat{K}_{\pm}]=\pm\hat{K}_{\pm}\quad,\quad[\hat{K}_{+},\hat{K}_{-}]=-2\hat{K}_{3}
\end{align}We then consider the following operator ansatz:
\begin{align}
\nonumber
\hat{F}_{1}(\theta)&=\!\exp\big(\theta\gamma_{+}\hat{K}_{+}\!+\!\theta\gamma_{-}\hat{K}_{-}\!+\!\theta\gamma_{3}\hat{K}_{3}\big)\\
\label{SupposeOPE1to3}
&=\!\exp\big(f_{+}(\theta)\hat{K}_{+}\big)\exp\big(f_{3}(\theta)\hat{K}_{3}\big)\exp\big(f_{-}(\theta)\hat{K}_{-}\big)
\end{align}where the coefficient functions $f_{\pm}(\theta)$ and $f_{3}(\theta)$ are to be determined. These functions can be obtained through the procedure outlined below.
Taking the derivative with respect to $\theta$, the first line of the identity \eqref{SupposeOPE1to3} yields
\begin{small}
\begin{align}
\nonumber
\frac{d\hat{F}_{1}(\theta)}{d\theta}&=\!(\gamma_{+}\hat{K}_{+}\!+\!\gamma_{-}\hat{K}_{-}\!+\!\gamma_{3}\hat{K}_{3})\cdot\text{e}^{(\theta\gamma_{+}\hat{K}_{+}+\theta\gamma_{-}\hat{K}_{-}+\theta\gamma_{3}\hat{K}_{3})}\\
\label{LHsDF1Dtheta}
&=\big(\gamma_{+}\hat{K}_{+}\!+\!\gamma_{-}\hat{K}_{-}\!+\!\gamma_{3}\hat{K}_{3}\big)\cdot\hat{F}_{1}(\theta)
\end{align}
\end{small}In deriving \eqref{LHsDF1Dtheta}, it is necessary to employ the differential identity for the exponential map,
\begin{align}
&\frac{d}{dt}\text{e}^{\hat{X}(t)}\!=\!\text{e}^{\hat{X}(t)}\frac{1-\text{e}^{-\text{ad}_{\hat{X}}}}{\text{ad}_{\hat{X}}}\frac{d\hat{X}(t)}{dt}\!=\!\text{e}^{\hat{X}(t)}\sum_{k=0}^{\infty}\frac{(-1)^{k}}{(k+1)!}(\text{ad}_{\hat{X}})^{k}\frac{d\hat{X}(t)}{dt},~ \text{ad}_{\hat{X}}\hat{Y}\!=\![\hat{X},\hat{Y}]
\end{align}which satisfies the Leibniz rule.
On the other hand, differentiating the second line of \eqref{SupposeOPE1to3} leads to
\begin{align}
\nonumber
&\frac{d\hat{F}_{1}(\theta)}{d\theta}\!=\!\frac{d}{d\theta}\big(\text{e}^{f_{+}(\theta)\hat{K}_{+}}\text{e}^{f_{3}(\theta)\hat{K}_{3}}\text{e}^{f_{-}(\theta)\hat{K}_{-}}\big)\!=\!f_{+}^{\prime}(\theta)\hat{K}_{+}\cdot\text{e}^{f_{+}(\theta)\hat{K}_{+}}\text{e}^{f_{3}(\theta)\hat{K}_{3}}\text{e}^{f_{-}(\theta)\hat{K}_{-}}\\
\nonumber
&\quad\quad\quad+\text{e}^{f_{+}(\theta)\hat{K}_{+}}f_{3}^{\prime}(\theta)\hat{K}_{3}\text{e}^{-f_{+}(\theta)\hat{K}_{+}}\cdot\text{e}^{f_{+}(\theta)\hat{K}_{+}}\text{e}^{f_{3}(\theta)\hat{K}_{3}}\text{e}^{f_{-}(\theta)\hat{K}_{-}}\\
\label{RHsDF1Dtheta}
&\quad\quad\quad+\text{e}^{f_{+}(\theta)\hat{K}_{+}}\text{e}^{f_{3}(\theta)\hat{K}_{3}}f_{-}^{\prime}(\theta)\hat{K}_{-}\text{e}^{-f_{3}(\theta)\hat{K}_{3}}\text{e}^{-f_{+}(\theta)\hat{K}_{+}}\cdot\text{e}^{f_{+}(\theta)\hat{K}_{+}}\text{e}^{f_{3}(\theta)\hat{K}_{3}}\text{e}^{f_{-}(\theta)\hat{K}_{-}}
\end{align}By further applying the identity
\begin{align}
&\text{e}^{\hat{X}}\hat{Y}\text{e}^{-\hat{X}}=\hat{Y}+[\hat{X},\hat{Y}]+\frac{1}{2!}[\hat{X},[\hat{X},\hat{Y}]]+\frac{1}{3!}[\hat{X},[\hat{X},[\hat{X},\hat{Y}]]]+\dots
\end{align}together with the commutation relations in \eqref{liealgebrasu11}, one can obtain
\begin{align}
\label{DoubleK+OnK3}
&\text{e}^{f_{+}(\theta)\hat{K}_{+}}\hat{K}_{3}\text{e}^{-f_{+}(\theta)\hat{K}_{+}}=\hat{K}_{3}-f_{+}(\theta)\hat{K}_{+}\\
\label{DoubleK+K3OnK-}
&\text{e}^{f_{+}(\theta)\hat{K}_{+}}\text{e}^{f_{3}(\theta)\hat{K}_{3}}\hat{K}_{-}\text{e}^{-f_{3}(\theta)\hat{K}_{3}}\text{e}^{-f_{+}(\theta)\hat{K}_{+}}=\text{e}^{-f_{3}(\theta)}\big(\hat{K}_{-}-2f_{+}(\theta)\hat{K}_{3}+f_{+}(\theta)^{2}\hat{K}_{+}\big)
\end{align}Substituting $\eqref{DoubleK+OnK3}$ and $\eqref{DoubleK+K3OnK-}$ into $\eqref{RHsDF1Dtheta}$, we obtain
\begin{align}
\label{RHsDF1DthetaV1}
&\frac{d\hat{F}_{1}(\theta)}{d\theta}\!=\!\big((f_{+}^{\prime}\!-\!f_{3}^{\prime}f_{+}\!+\!\text{e}^{-f_{3}}f_{-}^{\prime}f_{+}^{2})\hat{K}_{+}\!+\!\text{e}^{-f_{3}}f_{-}^{\prime}\hat{K}_{-}\!+\!(f_{3}^{\prime}\!-\!2\text{e}^{-f_{3}}f_{-}^{\prime}f_{+})\hat{K}_{3}\big)\cdot\hat{F}_{1}(\theta)
\end{align}By matching $\eqref{LHsDF1Dtheta}$ with $\eqref{RHsDF1DthetaV1}$, one derives a set of differential equations that determine the functions $f_{+}(\theta)$, $f_{-}(\theta)$, and $f_{3}(\theta)$:
\begin{align}
&\gamma_{+}=f_{+}^{\prime}-f_{3}^{\prime}f_{+}+\text{e}^{-f_{3}}f_{-}^{\prime}f_{+}^{2}\\
&\gamma_{3}=f_{3}^{\prime}-2\text{e}^{-f_{3}}f_{-}^{\prime}f_{+}\\
&\gamma_{-}=\text{e}^{-f_{3}}f_{-}^{\prime}
\end{align}From the ansatz \eqref{SupposeOPE1to3}, the boundary conditions at $\theta=0$ are readily identified as $f_{+}(0)=f_{-}(0)=f_{3}(0)=0$. Solving the above system yields
\begin{align}
&\begin{cases}
\begin{array}{c}
f_{+}(\theta)=\frac{\sqrt{4\gamma_{+}\gamma_{-}-\gamma_{3}^{2}}\tan\big(\frac{1}{2}\sqrt{4\gamma_{+}\gamma_{-}-\gamma_{3}^{2}}(\theta+\frac{\text{arctanh}(-\gamma_{3}/2\beta)}{\beta})\big)-\gamma_{3}}{2\gamma_{-}}\\
f_{-}(\theta)=-\frac{\gamma_{3}}{2\gamma_{+}}-\frac{2\beta^{2}\tan\big(\frac{1}{2}\sqrt{4\gamma_{+}\gamma_{-}-\gamma_{3}^{2}}(\theta+\frac{\text{arctanh}(-\gamma_{3}/2\beta)}{\beta})\big)}{\gamma_{+}\sqrt{4\gamma_{+}\gamma_{-}-\gamma_{3}^{2}}}\\
f_{3}(\theta)=-\log\big(\frac{-\gamma_{+}\gamma_{-}}{\beta^{2}}\big)-2\log\big(\cos\big(\frac{1}{2}\sqrt{4\gamma_{+}\gamma_{-}-\gamma_{3}^{2}}(\theta+\frac{\text{arctanh}(-\gamma_{3}/2\beta)}{\beta})\big)\big)
\end{array}\end{cases}
\end{align}
where $\beta^{2}=\frac{\gamma_{3}^{2}}{4}-\gamma_{+}\gamma_{-}$. Setting $\theta=1$, the operator expansion in $\eqref{SupposeOPE1to3}$ reduces to
\begin{align}
\label{Soltheta1OPE1to3}
&\exp\big(\gamma_{+}\hat{K}_{+}+\gamma_{-}\hat{K}_{-}+\gamma_{3}\hat{K}_{3}\big)=\exp(\Gamma_{+}\hat{K}_{+})\exp(\Gamma_{3}\hat{K}_{3})\exp(\Gamma_{-}\hat{K}_{-})
\end{align}
with the coefficients given by
\begin{small}
\begin{align}
&\Gamma_{+}\!=\!\frac{2\tanh(\beta)\gamma_{+}}{2\beta-\gamma_{3}\tanh(\beta)},\Gamma_{-}\!=\!\frac{2\tanh(\beta)\gamma_{-}}{2\beta-\gamma_{3}\tanh(\beta)},\Gamma_{3}\!=\!-2\log\big(\cosh(\beta)-\frac{\gamma_{3}}{2\beta}\sinh(\beta)\big)
\end{align}
\end{small}

The identity \eqref{Soltheta1OPE1to3} provides a general disentangling formula for the Lie algebra $su(1,1)$. It remains valid even when the coefficients $\gamma_{\pm,3}$ and the generators $\hat{K}_{\pm,3}$ carry additional indices, provided the corresponding summations are appropriately included. Returning to the present setup, we express the generators of $su(1,1)$ in terms of creation and annihilation operators, as given in \eqref{KplusCreationAnnihilation}–\eqref{KminusCreationAnnihilation}. The squeezing operator $\hat{\mathcal{S}}\big(\gamma(\tau),\varphi(\tau)\big)$ can then be rewritten as
\begin{align}
&\hat{\mathcal{S}}^{(\mathtt{i})}(\gamma,\varphi)=\exp\bigg(\frac{1}{2\pi}\sum_{m=2}\big\{\underbrace{(-\text{i}\gamma_{m}^{(\mathtt{i})}\text{e}^{2\text{i}\varphi_{m}^{(\mathtt{i})}})}_{\gamma_{m,+}^{(\mathtt{i})}(\tau)}\hat{\mathcal{K}}_{m,+}^{(\mathtt{i})}(\tau_{0})+\underbrace{(-\text{i}\gamma_{m}^{(\mathtt{i})}\text{e}^{-2\text{i}\varphi_{m}^{(\mathtt{i})}})}_{\gamma_{m,-}^{(\mathtt{i})}(\tau)}\hat{\mathcal{K}}_{m,-}^{(\mathtt{i})}(\tau_{0})\big\}\bigg)
\end{align}where the weight-shift operators $\hat{\mathcal{K}}{m,+}^{(\mathtt{i})}(\tau{0})$ and $\hat{\mathcal{K}}{m,-}^{(\mathtt{i})}(\tau{0})$ satisfy the $su(1,1)$ algebra, as discussed in Appendix~\ref{AppendixNestedCommuta}. We emphasize that no summation is implied over the polarization index $(\mathtt{i})$. By applying the disentangling formula \eqref{Soltheta1OPE1to3} to the above expression, the squeezing operator $\hat{\mathcal{S}}^{(\mathtt{i})}(\gamma,\varphi)$ can be factorized as
\begin{align}
\nonumber
&\hat{\mathcal{S}}^{(\mathtt{i})}(\gamma,\varphi)=\exp\big\{\frac{1}{2\pi}\sum_{m=2}\big(\text{e}^{2\text{i}\varphi_{m}^{(\mathtt{i})}(\tau)}\tanh(\gamma_{m}^{(\mathtt{i})}(\tau))\hat{a}_{m}^{(\mathtt{i})\dagger}(\tau_{0})\hat{a}_{-m}^{(\mathtt{i})\dagger}(\tau_{0})\big)\big\}\\
\nonumber
&\cdot\exp\big\{\frac{1}{2\pi}\sum_{m=2}\big(-\ln(\cosh(\gamma_{m}^{(\mathtt{i})}(\tau)))\big(\hat{a}_{-m}^{(\mathtt{i})}(\tau_{0})\hat{a}_{-m}^{(\mathtt{i})\dagger}(\tau_{0})+\hat{a}_{m}^{(\mathtt{i})\dagger}(\tau_{0})\hat{a}_{m}^{(\mathtt{i})}(\tau_{0})\big)\big)\big\}\\
&\cdot\exp\big\{\frac{1}{2\pi}\sum_{m=2}\big(-\text{e}^{-2\text{i}\varphi_{m}^{(\mathtt{i})}(\tau)}\tanh(\gamma_{m}^{(\mathtt{i})}(\tau))\hat{a}_{-m}^{(\mathtt{i})}(\tau_{0})\hat{a}_{m}^{(\mathtt{i})}(\tau_{0})\big)\big\}
\end{align}

\bibliographystyle{unsrt}
\bibliography{bibliography}

\end{document}